\documentclass[twocolumn]{aastex701}

\usepackage{amsmath}
\usepackage{amssymb}

\newcommand{\h}{\ensuremath{h_0}}
\newcommand{\hR}{\ensuremath{h_0/R_d}}

\begin{document}

\title{Through Thick and Thin: The Cosmic Evolution of Disk Scale Height}

\author[orcid=0000-0002-3462-4175]{Si-Yue Yu}
\affiliation{Department of Astronomy, Xiamen University, Xiamen, Fujian 361005, People's Republic of China}
\affiliation{Kavli Institute for the Physics and Mathematics of the Universe (WPI), The University of Tokyo Institutes for Advanced Study, The University of Tokyo, Kashiwa, Chiba 277-8583, Japan}
\email[show]{syu@xmu.edu.cn}

\author[orcid=0000-0001-6947-5846]{Luis C. Ho}
\affiliation{The Kavli Institute for Astronomy and Astrophysics, Peking University, 5 Yiheyuan Road, Haidian District, Beijing, 100871, China}
\affiliation{Department of Astronomy, Peking University, 5 Yiheyuan Road, Haidian District, Beijing, 100871, China}
\email{lho.pku@gmail.com}

\author[orcid=0000-0002-1499-6377]{Takafumi Tsukui}
\affiliation{Kavli Institute for the Physics and Mathematics of the Universe (WPI), The University of Tokyo Institutes for Advanced Study, The University of Tokyo, Kashiwa, Chiba 277-8583, Japan}
\email{takafumi.tsukui@ipmu.jp}

\author[0000-0002-0000-6977]{John D. Silverman}
\affiliation{Kavli Institute for the Physics and Mathematics of the Universe (WPI), The University of Tokyo Institutes for Advanced Study, The University of Tokyo, Kashiwa, Chiba 277-8583, Japan}
\affiliation{Department of Astronomy, School of Science, The University of Tokyo, 7-3-1 Hongo, Bunkyo, Tokyo 113-0033, Japan}
\affiliation{Center for Data-Driven Discovery, Kavli IPMU (WPI), UTIAS, The University of Tokyo, Kashiwa, Chiba 277-8583, Japan}
\affiliation{Center for Astrophysical Sciences, Department of Physics \& Astronomy, Johns Hopkins University, Baltimore, MD 21218, USA}
\email{john.silverman@ipmu.jp}

\author[orcid=0000-0002-1416-8483]{Marc Huertas-Company}
\affiliation{Instituto de Astrofísica de Canarias (IAC), La Laguna, E-38205, Spain}
\affiliation{Observatoire de Paris, LERMA, PSL University, 61 avenue de l’Observatoire, F-75014 Paris, France}
\affiliation{Université Paris-Cité, 5 Rue Thomas Mann, 75014 Paris, France}
\affiliation{Universidad de La Laguna. Avda. Astrofísico Fco. Sanchez, La Laguna, Tenerife, Spain}
\affiliation{Center for Computational Astrophysics, Flatiron Institute, New York, NY 10010, USA}
\email{marc.huertas.company@gmail.com}

\author[orcid=0000-0002-6610-2048]{Anton M. Koekemoer}
\affiliation{Space Telescope Science Institute, 3700 San Martin Drive, Baltimore, MD 21218, USA}
\email{koekemoer@stsci.edu}

\author[0000-0002-3560-8599]{Maximilien Franco}
\affiliation{Université Paris-Saclay, Université Paris Cité, CEA, CNRS, AIM, 91191 Gif-sur-Yvette, France}
\email{maximilien.franco@cea.fr}

\author[0000-0002-6085-3780]{Richard Massey}
\affiliation{Department of Physics, Centre for Extragalactic Astronomy, Durham University, South Road, Durham DH1 3LE, UK}
\email{r.j.massey@durham.ac.uk}

\author[0000-0002-8434-880X]{Lilan Yang}
\affiliation{Laboratory for Multiwavelength Astrophysics, School of Physics and Astronomy, Rochester Institute of Technology, 84 Lomb Memorial Drive, Rochester, NY 14623, USA}
\email{lxysps@rit.edu}

\author[0000-0002-0569-5222]{Rafael C. Arango-Toro}
\affiliation{Aix Marseille Univ, CNRS, CNES, LAM, Marseille, France  }
\email{rafael.arango-toro@lam.fr}

\author[0000-0002-9382-9832]{Andreas L. Faisst}
\affiliation{Caltech/IPAC, MS 314-6, 1200 E. California Blvd. Pasadena, CA 91125, USA}
\email{afaisst@caltech.edu}

\author[0000-0002-0236-919X]{Ghassem Gozaliasl}
\affiliation{Department of Computer Science, Aalto University, P.O. Box 15400, FI-00076 Espoo, Finland}
\affiliation{Department of Physics, University of, P.O. Box 64, FI-00014 Helsinki, Finland}
\email{ghassem.gozaliasl@gmail.com}

\author[0000-0002-5496-4118]{Kartik Sheth}
\affiliation{Empowered Earth Alliance, Washington, DC 20003, USA}
\email{astrokartik@gmail.com}

\author[0000-0001-9187-3605]{Jeyhan S. Kartaltepe}
\affiliation{Laboratory for Multiwavelength Astrophysics, School of Physics and Astronomy, Rochester Institute of Technology, 84 Lomb Memorial Drive, Rochester, NY 14623, USA}
\email{jeyhan@astro.rit.edu}

\author[0000-0002-8437-6659]{Can Xu}
\affiliation{School of Astronomy and Space Science, Nanjing University, Nanjing, Jiangsu 210093, China}
\affiliation{Key Laboratory of Modern Astronomy and Astrophysics, Nanjing University, Ministry of Education, Nanjing 210093, China}
\affiliation{Kavli Institute for the Physics and Mathematics of the Universe (WPI), The University of Tokyo Institutes for Advanced Study, The University of Tokyo, Kashiwa, Chiba 277-8583, Japan}
\email{canxu@smail.nju.edu.cn}

\author[0009-0006-3071-7143]{Aryana Haghjoo}
\affiliation{Department of Physics and Astronomy, University of California, Riverside, 900 University Ave, Riverside, CA 92521, USA}
\email{aryana.haghjoo@email.ucr.edu}

\author[0000-0001-8917-2148]{Xuheng Ding}
\affiliation{School of Physics and Technology, Wuhan University, Wuhan 430072, China}
\affiliation{Kavli Institute for the Physics and Mathematics of the Universe (WPI), The University of Tokyo Institutes for Advanced Study, The University of Tokyo, Kashiwa, Chiba 277-8583, Japan}
\email{dingxh@whu.edu.cn}

\author[0000-0002-9252-114X]{Zhaoxuan Liu}
\affiliation{Kavli Institute for the Physics and Mathematics of the Universe (WPI), The University of Tokyo Institutes for Advanced Study, The University of Tokyo, Kashiwa, Chiba 277-8583, Japan}
\affiliation{Center for Data-Driven Discovery, Kavli IPMU (WPI), UTIAS, The University of Tokyo, Kashiwa, Chiba 277-8583, Japan}
\affiliation{Department of Astronomy, School of Science, The University of Tokyo, 7-3-1 Hongo, Bunkyo, Tokyo 113-0033, Japan}
\email{zhaoxuan.liu@ipmu.jp}

\author[0000-0002-9883-7460]{Jacqueline E. McCleary}
\affiliation{Department of Physics, Northeastern University, 360 Huntington Ave, Boston, MA}
\email{j.mccleary@northeastern.edu}

\begin{abstract}
To investigate the formation and evolution of vertical structures in disk galaxies, we measure global $\operatorname{sech}^2$ scale heights, averaging thin and thick components when present, for 2631 edge-on disk galaxies with $M_*>10^{10}\,M_\odot$ at $0<z<3.5$ from the JWST COSMOS-Web survey. We show that dust extinction systematically overestimates scale heights at shorter rest-frame wavelengths, and therefore adopt a fixed rest-frame wavelength of 1\,\micron. After further correcting for projection-induced bias using a new accurate method, we find that the median disk scale height increases from $0.56\pm0.03$\,kpc at $z=3.25$ to $0.84\pm0.04$\,kpc at $z=1.25$, and subsequently decreases to $0.67\pm0.06$\,kpc at $z=0.25$. The bias-corrected disk scale-length-to-height ratio remains constant at $2.7\pm0.2$ for $z>1.5$, but rises to $4.0\pm0.4$ at $z=0.25$. These results imply that the high-redshift progenitors of present-day thick disks were of intermediate thickness, neither thin nor thick, yet dynamically hot and dense. The observed radial variation of scale height is consistent with the artificial flaring expected from observational effects, disfavoring minor mergers as the primary mechanism of disk thickening. Instead, we suggest that the high-redshift intermediate-thickness disks were single-component systems that increased their vertical scale height through decreasing surface mass density and/or violent gravitational instabilities, eventually producing thick disks. Thin-disk growth begins at $z\approx2$ and dominates at $z\lesssim1$, yielding a vertically more compact system with decreasing scale heights from $z\approx1$ to $0$. The inferred thin-disk mass fraction increases from $0.1\pm0.03$ at $z=1$ to $0.6\pm0.1$ at $z=0$. Together, these findings reveal a continuous evolutionary link between high-redshift single-component disks and present-day thick thin disk systems.
\end{abstract}

\keywords{\uat{High-redshift galaxies}{734} --- \uat{Galaxy kinematics}{602} --- \uat{Galaxy photometry}{611} --- \uat{Galaxy stellar disks}{1594} --- \uat{Galaxy structure}{622} --- \uat{Milky Way evolution}{1052} --- \uat{Galaxy evolution}{594}}

\section{Introduction} \label{intro}
The thick disk was first identified in 1979 as a diffuse stellar component in edge-on S0 galaxies, distinct from both the bulge and thin disk \citep{Burstein1979}, and was subsequently recognized in the Milky Way in 1983 through star counts toward the South Galactic Pole \citep{GilmoreReid1983}. These discoveries established the now-standard view that the Milky Way disk comprises two components: a thin and a thick disk \citep{FreemanBland-Hawthorn2002, Bensby2014, Bland-Hawthorn2016}.

The formation history of the Milky Way’s disk appears to have proceeded in two nearly disjoint phases \citep{Nissen2020, Xiang2022, xiang2025}. The thin disk, with an exponential scale height of $\sim$0.3--0.4\,kpc \citep{Juric2008, Bland-Hawthorn2016}, is dominated by younger stars and was formed from the cold gaseous mid-plane during the phase of steady star formation over the past $\sim$8\,Gyr \citep{Forbes2012, Xiang2022, YuSijie2023}. The thick disk, with an exponential scale height of $\sim$0.7--1.2\,kpc, consists primarily of old, [$\alpha$/Fe]-enhanced stars that originated during an earlier, bursty star formation phase within the first $\sim$5\,Gyr \citep{Juric2008, Bensby2014, Bland-Hawthorn2016, YuSijie2021, YuSijie2023}. Thick disks are now known to be nearly ubiquitous among nearby disk galaxies \citep{YoachimDalcanton2006, Bizyaev2009, Bizyaev2014, Comeron2014, Comeron2016, Comeron2018, Kauffmann2025}. 

Despite their prevalence, the physical origin of thick disks remains debated. They may have been born as inherently thick structures under turbulent, gas-rich conditions in the early Universe \citep{Bournaud2009, YuSijie2023}. Alternative, they may have grown from initially thin disks through heating by minor mergers or internal secular processes \citep{Quinn1993, Villalobos2008, Qu2011, Bird2021}.

Studying galaxy disks across cosmic time provides crucial insight into the mechanisms that construct their vertical structure. In the HST era, the limited spatial resolution made it difficult to distinguish thin and thick disks, restricting studies to global scale-height measurements. Tracing rest-frame optical to UV wavelength, \citet{Elmegreen2017} reported a median global scale height of $0.63$\,kpc, with a scatter of $0.24\,$kpc among individual measurements, at $z\sim2$ (see also \citealt{Elmegreen2006}) and detected vertical color gradients that suggest the coexistence of thin and thick components at early epochs. A later analysis by \citet{HamiltonCampos2023} yielded a median scale height of $0.74$\,kpc, with a scatter of $0.35\,$kpc, after correcting for the overestimation caused by projection effect of deviations from perfect edge-on inclination; however, such corrections depend essentially on the ratio of scale height to scale length, a dependence explored in this study.

Progress in characterizing disk structure beyond $z\sim1$ has recently accelerated thanks to the unprecedented sensitivity and resolution of JWST's Near-Infrared Camera (NIRCam). JWST observations have revealed a substantial population of regular disk galaxies already in place within the first few Gyrs, suggesting that disk assembly and evolution began earlier than previously thought \citep{Ferreira2022b, Kartaltepe2023, Nelson2023, Robertson2023, Jacobs2023, Cheng2022b, Cheng2023, LeConte2024, XuYu2024, Yu2025, Huertas-Company2025}. Using F115W imaging that traces rest-frame NIR at $z\sim0$ but UV at $z>3$ for a sample with a median stellar mass of $10^9\,M_\odot$, \citet{Lian2024} measured a median global scale height of $0.38$\,kpc, with a scatter of $0.13$\,kpc, at $z=0.2$--5, lower than the previous HST results.  \citet{Tsukui2025}\footnote{Due to different definitions of scale height (their Equation~(1)), the values reported by \citet{Tsukui2025} should be multiplied by a factor of 2 to match those presented here.}, using F277W, F356W, and F444W filter to trace rest-frame 1--2\,\micron\ wavelength at $z\leq 3$, obtained a median global scale height of $0.66\pm0.07$\,kpc (where all quoted values in this work follow the +/$-$ notation for uncertainties) for stellar mass of $10^{9.2}\,M_\odot$, consistent with earlier HST results. Moreover, they performed the first thick-thin disk decomposition at $z>1$, finding median scale heights of $0.33\pm0.04$\,kpc and $0.94\pm0.07$\,kpc for the thin and thick components, respectively.

Still, as these pioneering JWST studies do not adopt a fixed rest-frame wavelength for the measurements, the derived scale heights may be biased by extinction from mid-plane dust lanes or by spatial variations in stellar populations. Furthermore, their analyses were based on relatively small samples (fewer than 200 galaxies). Expanding to larger, statistically robust samples is essential to establish the redshift evolution of disk scale heights. Equally important is correcting for the bias introduced by variation in rest-frame wavelength and developing a more accurate method to correct for the bias introduced by deviations from a perfectly edge-on inclination. Motivated by these needs, we present a comprehensive analysis of edge-on disk galaxies observed by the COSMOS-Web survey \citep{Casey2023}. In this work, we focus on the global disk scale height, averaging thin and thick components when present, while future studies will address thin–thick disk decompositions.

This paper is organized as follows. Section~\ref{sec:sample} describes the sample selection and its properties. Section~\ref{sec:methods} details the image analysis and scale height measurements. Section~\ref{sec:results} presents the main results. Section~\ref{sec:implication} discusses the implications for the formation and evolution of galaxy disks. Finally, Section~\ref{sec:summary} summarizes the main findings. 
Throughout this paper, we adopt AB magnitudes, \cite{Chabrier2003} initial mass function, and assume a flat $\Lambda$CDM cosmology with $(\Omega_{\rm m}, \Omega_{\Lambda}) = (0.27, 0.73)$ and a Hubble constant of $H_0 = 70\,{\rm km\,s^{-1}\,Mpc^{-1}}$.

\begin{figure}
  \centering
  \includegraphics[width=\columnwidth]{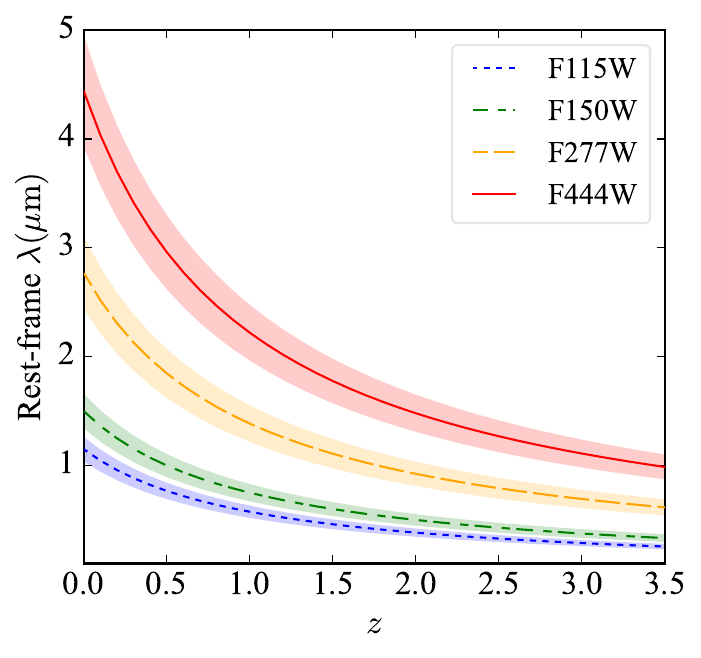}
  \caption{Observed rest-frame wavelengths of the NIRCam filters used in COSMOS-Web as a function of redshift. The shaded region marks the bandwidth of the filter. The upper redshift limit is chosen to ensure that the pivot wavelength of the F444W filter probes rest-frame wavelengths of at least 1\,$\micron$. We obtain the scale heights at a fixed rest-frame 1\,\micron\ by interpolating the measurements across the available filters (see Section~\ref{sect:rest}). 
  }
  \label{fig:rest}
\end{figure}

\begin{figure}
  \centering
  \includegraphics[width=\columnwidth]{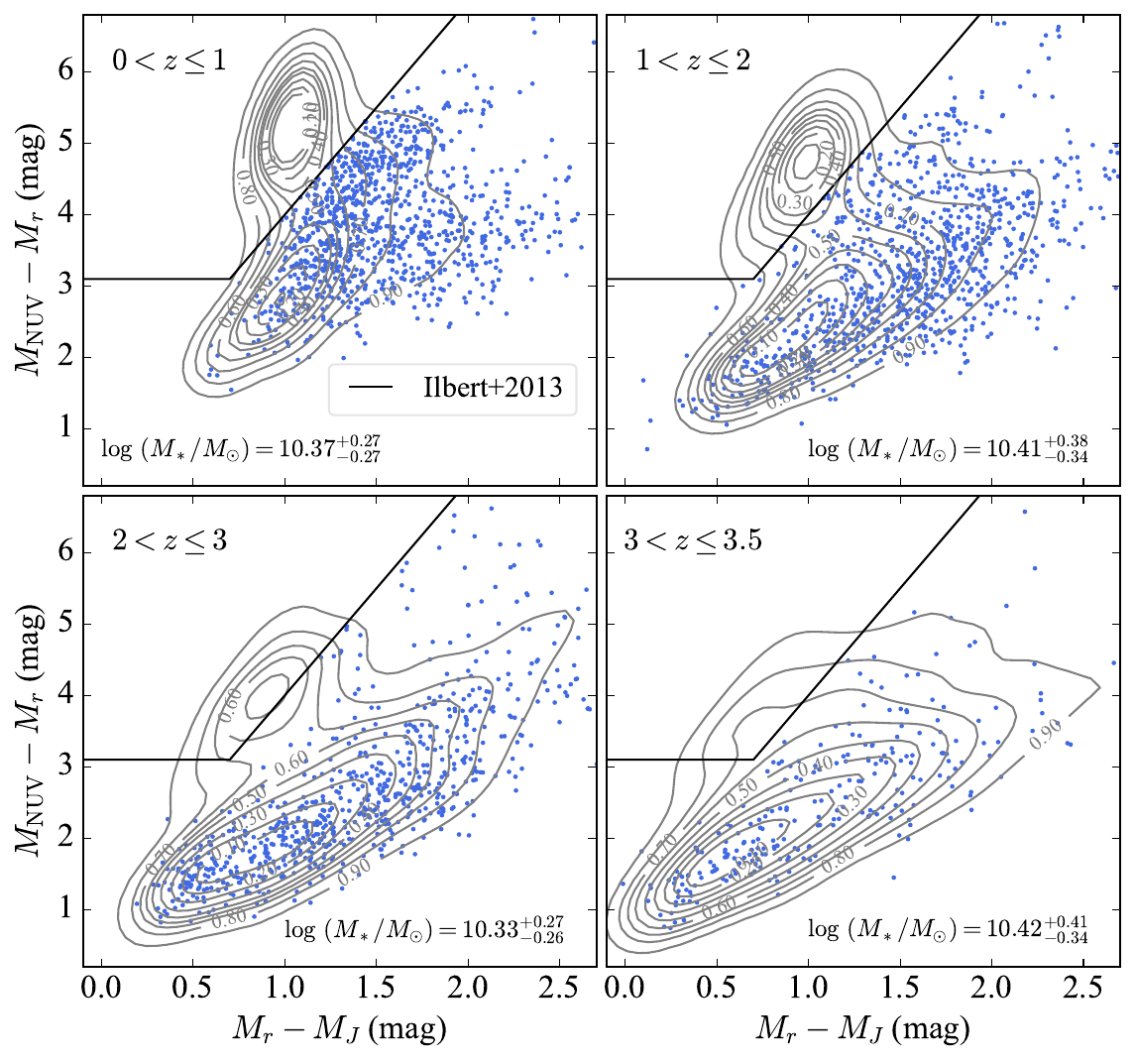}
  \caption{Diagram  of $M_r-M_J$ versus $M_{\rm NUV}-M_r$ for our edge-on disk galaxy sample (blue dots). Contours indicate regions enclosing a given fraction of galaxies in the parent sample within each redshift range.  
Quiescent galaxies, which mostly lie above the dividing line proposed by \citet{Ilbert2013}, are excluded.
  Median stellar mass and scatter for our sample for each redshift bin are shown at the bottom. 
   }
  \label{fig:color}
\end{figure}

\section{Sample and Data} \label{sec:sample}

The COSMOS-Web survey is the largest galaxy survey conducted in JWST Cycle-1 \citep{Casey2023}. Its NIRCam observations in four filters (F115W, F150W, F277W, and F444W) provide a deep and high-resolution dataset for our structural and photometric analysis. For this study, we use the first data release of COSMOS-Web NIRCam mosaics\footnote{ \url{https://cosmos2025.iap.fr} } \citep{Franco2025}. The mosaics have a pixel scale of $0\farcs03$ at all filters. \citet{Shuntov2025}  constructed the COSMOS-Web photometric catalog, by performing source extraction across 33 photometric bands spanning 0.3--8\,$\micron$ and measured photometry across available bands using a multi-band model-fitting approach with {\tt SourceX-Tractor++} \citep{Bertin2020, Kummel2022}. Using these measurements, \citet{Shuntov2025} derived physical properties for each source through template spectral energy distribution fitting using \texttt{LePhare} \citep{Arnouts2002, Ilbert2006}. From the COSMOS-Web catalog (version 1.1), we obtain photometric redshifts ($z$), stellar masses ($M_*$), and absolute magnitudes in the Near-Ultraviolet (NUV), $r$-, and $J$-bands ($M_{\rm NUV}$, $M_r$, and $M_J$). Figure~\ref{fig:rest} illustrates the rest-frame wavelength probed by each NIRCam filter as a function of redshift. In particular, the F444W filter observes rest-frame 1\,$\micron$ at $z=3.5$, above which the rest-frame pivot wavelength drops below 1\,$\micron$. To minimize the impact of dust extinction on our morphological measurements, we require that the reddest filter probes at least 1\,$\micron$. We therefore limit our analysis to galaxies within $0 < z < 3.5$. The choice of the upper redshift limit also aligns with the epoch of the Milky Way's thick disk formation \citep{Conroy2022}.

We further apply the following selection criteria to construct our parent sample. To ensure a clean and reliable galaxy sample, we require the following flags provided in the COSMOS-Web catalog: {\it type = $0$} (select galaxies instead of stars), {\it flag\_star\_hsc = $0$} (exclude sources severely contaminated by nearby stars), and {\it warn\_flag = $0$} (remove spurious detections such as hot pixels, sources detected in only a single filter, or other imaging artifacts). We restrict our analysis to galaxies with $M_*\geq10^{10}\,M_{\odot}$, as recent studies suggest that lower-mass galaxies at $z>1$ may be prolate rather than disky systems \citep{Pandya2024, Vega-Ferrero2024, Klein2025}. Applying these criteria yields a parent sample of 19199 galaxies.

We visually inspected each galaxy in the parent sample and excluded 1445 objects that were under interaction/merger or were too close to other galaxies or bright stars for reliable morphology measurements. This step results in a cleaned sample of 17754 galaxies. For each object, we use the Python package {\tt SEP} \citep{Bertin1996, Barbary2016} on the F444W images to generate a target galaxy segmentation map and generate a mask file for removing contaminating sources. Mask generation was performed with both cold and hot detection modes, with parameters adjusted manually. These masks are used for subsequent model fitting.

Figure~\ref{fig:color} shows $M_{\rm NUV}-M_r$ vs. $M_r-M_J$ color-color diagrams for four redshift bins. The background contours indicate the distribution of the parent sample, with contours enclosing 10\%, 20\%, ..., 90\% of galaxies. Quenched galaxies were identified using the division line (black line in Figure~\ref{fig:color}) proposed by \citet{Ilbert2013} and were excluded, leaving 13453 star-forming galaxies for analysis.  We first perform single Sérsic fitting using the code {\tt IMFIT} \citep{Erwin2015} and remove unresolved sources with effective radius in F444W band less than F444W PSF FWHM. We then perform bulge-disk decomposition, modeling each galaxy with an exponential disk and a round Sérsic bulge with the Sérsic index fixed at 4, in all available filters. The generation of point-spread functions (PSFs) used in the fitting is described in Section~\ref{hybrid}. Catastrophic fitting results are excluded. From this decomposition, we calculate the bulge-to-total light ratio ($B/T$) and exclude potential elliptical galaxies, either introduced by photometric measurement uncertainties or representing star-forming ellipticals, by adopting a threshold of $B/T>0.5$. Using a higher threshold (e.g., $B/T=0.7$) does not affect our main results (see Section~\ref{sec:bias_remove}). We measure the axis ratio of the disk component ($q = b/a$, where $a$ and $b$ are the semi-major and semi-minor axes, respectively) in each filter. Using interpolation, we derive the disk axis ratio at a consistent rest-frame wavelength of 1\,$\micron$ for all galaxies, which we define as the final disk axis ratio ($q_d$). Edge-on disks appear as flattened systems and can be efficiently selected using $q_d$. This criterion is objective and reproducible, in contrast to visual classification. We therefore select our nearly edge-on disk sample by requiring $q_d < 0.4$, following the criterion adopted by \citet{HamiltonCampos2023}, which is slightly more permissive than the $q_d < 0.3$ threshold used by \citet{Elmegreen2017}. As demonstrated in Section~\ref{sect:incl}, our method for correcting biases introduced by projection effects is effective, such that adopting either $q_d < 0.3$ or $q_d < 0.4$ yields nearly identical bias-corrected scale heights. The above criteria reduce the 13453 galaxies to 2631 edge-on disks, the sample studied in this work. This sample is 15--30 times larger than in previous studies.

In contrast to our PSF-corrected measurements, applying an axis-ratio cut based on values derived without PSF correction (such as those from segmentation analysis), as done in some previous studies, can bias the sample toward intrinsically flatter systems at higher redshifts due to the smaller apparent sizes of galaxies. Our edge-on disk sample is shown in Figure~\ref{fig:color} in the color-color diagram (blue dots), with median stellar mass values ($\sim2.5\times10^{10}\,M_\odot$) and scatter indicated at the bottom. The stellar masses of our sample lie well above the completeness limits at each redshift \citep{Shuntov2025}.

We note that there are apparent clustering of galaxies at some specific redshift is a known artifact of the spectral template fitting process for determining the photometric redshift. This can occur when prominent spectral features, such as the Balmer or 4000\,\AA\ break, transition between photometric bands.  We have verified that the clustering is observed for both edge-on and face-on galaxies, indicating that the selection of edge-on disks does not introduce any unusual biases caused by the redshift determination. The overall quality of the photometric redshifts in our sample is high, with a catastrophic failure rate of only 1.44\% and a normalized median absolute deviation of 0.011 compared to spectroscopic redshifts \citep{Shuntov2025}.

\section{Measurements} \label{sec:methods}

\subsection{Hybrid PSFs}\label{hybrid}

\begin{figure}
  \centering
  \includegraphics[width=\columnwidth]{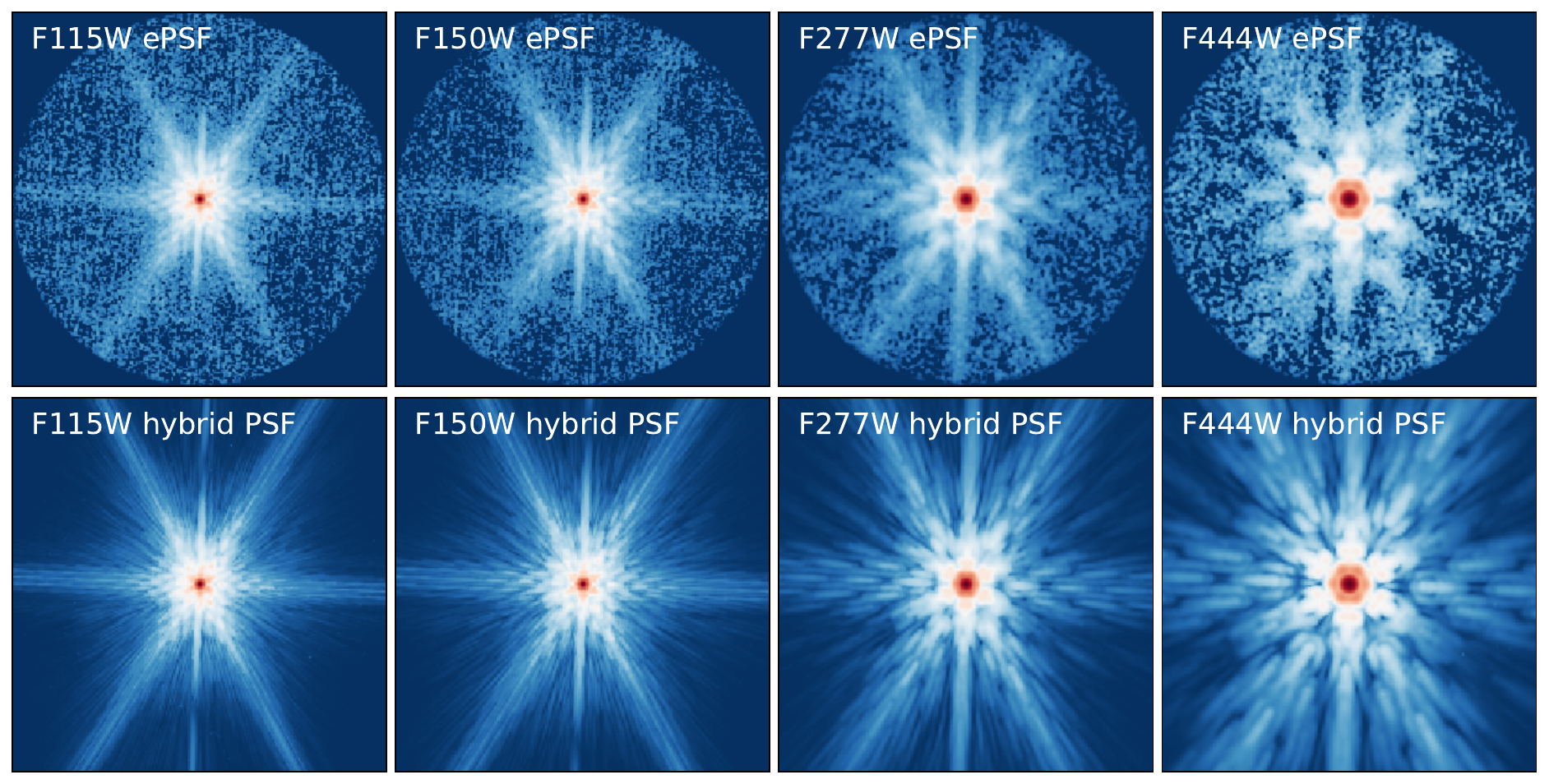}
  \caption{Comparison of the (top) ePSF and (bottom) our hybrid PSF for each JWST NIRCam filter of COSMOS-Web. The ePSF is constructed using PSFEX, while the hybrid PSF is constructed by replacing the low-$S/N$ pixels in ePSF with the values from simulated theoretical PSFs using a smooth top-hat function. These hybrid PSFs are adopted for all model fittings throughout this work.  }
  \label{fig:psf_comp}
\end{figure}

Accurate PSFs are essential for robust morphological fitting of galaxies. Theoretical PSFs simulated with {\tt WebbPSF} \citep{Perrin2014} tend to be narrower than those derived from real stars in drizzled mosaics \citep{Ono2023, Zhuang2024}, as {\tt WebbPSF} is designed to simulate the PSF on a single exposure, and should not be expected to agree with the PSF from drizzled mosaics, which include additional pixel-level convolutions.  Since our analysis is based on drizzled mosaics, we construct empirical PSFs (ePSFs) from field-star images in each NIRCam filter using {\tt PSFEx} \citep{PSFEx2011}, as shown in the top row of Figure~\ref{fig:psf_comp}. Still, these ePSFs suffer from relatively low signal-to-noise ratios ($S/N$) in their outer regions.  A common strategy to improve PSF quality is to combine the high-$S/N$ core of an empirical PSF with the noise-free wings of a theoretical one, forming a hybrid PSF \citep[e.g.,][]{vanderwel2012, Chen2025}. Following this approach, we generate theoretical PSFs for each NIRCam filter using {\tt WebbPSF}\footnote{ \url{https://stpsf.readthedocs.io/en/latest/} } and rotate them to match the orientation angle of the corresponding ePSFs by minimizing their difference. We then replace the noisy pixels (approximately the faintest 10\% in flux) of the ePSF with values from the theoretical PSF. A smoothed step function and reversed step function are applied during this combination to ensure a seamless transition. Finally, we normalize the combined image to produce the hybrid PSF, as illustrated in the bottom row of Figure~\ref{fig:psf_comp}.  These hybrid PSFs, which preserve the realistic core structure of observed stars while maintaining accurate outer profiles, are adopted for all model fittings throughout this work.

\begin{figure*}
  \centering
  \includegraphics[width=2\columnwidth]{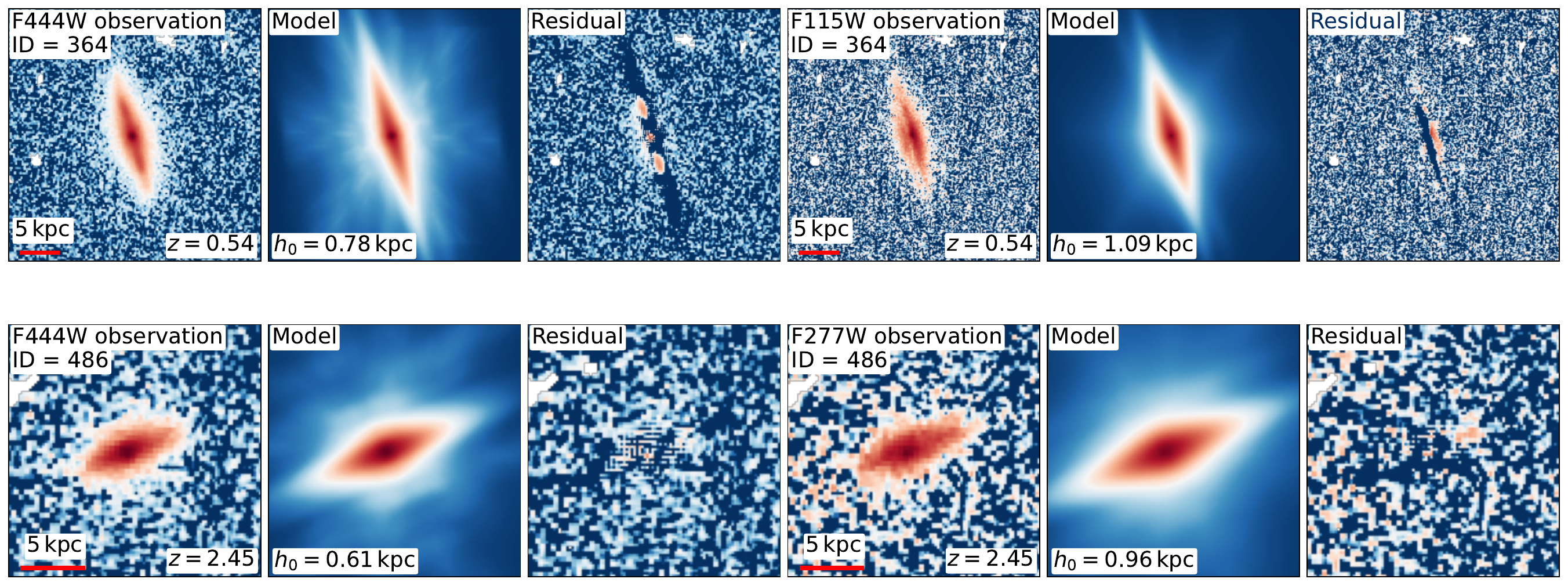}
  \caption{Example of the Sérsic-sech$^2$ model fitting. The top row shows a galaxy (ID\,=\,364) at $z=0.54$, with the left and right three panels displaying the results for the F444W and F115W filters, respectively. The bottom row shows a galaxy (ID\,=\,486) at $z=2.45$, with results for the F277W and F115W filters. A scale bar is shown at the bottom-left corner. The derived sech$^2$ vertical scale height ($h_0$) is indicated in each case.
  }
  \label{fig:example}
\end{figure*}

\subsection{Scale-height measurement} \label{h_measure}

The scale height provides a quantitative measure of disk thickness, with larger values corresponding to thicker disks. In a coordinate system aligned with the edge-on inclination, we define the radial distance along the disk plane (parallel to the major axis) as $R$, and the vertical distance from the plane (parallel to the minor axis) as $h$. The galaxy surface brightness at $(R, h)$ is denoted as $I(R,h)$. Assuming an isothermal disk, the light distribution of an edge-on disk can be described by \citep{Kruit1981}:
\begin{equation}\label{Eq:general}
  I(R,h) = I_0 \cdot (R/R_d) \cdot K_1(R/R_d) \cdot {\rm sech}^{2/m}(mh/h_0),
\end{equation}

\noindent
where $I_0$ is the central intensity, $K_1$ is the modified Bessel function of the second kind of order one, $R_d$ is the exponential scale length in the disk plane, $h_0$ is the scale height, and $m$ controls the overall shape of the vertical profile. Because $m$ and $h_0$ are degenerate, $m$ is usually fixed to 1 or $+\infty$ when determining the scale height, and  Equation~(\ref{Eq:general}) then simplifies to the following two common forms: a sech$^2$ function,
\begin{equation}\label{Eq:sech2_}
  I \propto {\rm sech}^{2}(h/h_0),
\end{equation}
\noindent
and an exponential function, 
\begin{equation}\label{Eq:exp_}
  I \propto \exp(-2h/h_0 ).
\end{equation}
\noindent
Due to the above functional forms, it is generally approximated that the sech$^2$ scale height  is twice the exponential scale height \citep[e.g.,][]{YoachimDalcanton2006}. However, this approximation is not exact, as the vertical profile of a galaxy is independent of the functional form adopted. \cite{Bland-Hawthorn2016} suggest the sech$^2$ scale height is 1.8 times the exponential scale height, whereas our analysis (see below) indicates that a factor of 1.37 provides a more accurate conversion for our sample.

Two-dimensional (2D) model fitting is an efficient approach for measuring scale heights \citep[e.g.,][]{YoachimDalcanton2006, Ranaivoharimina2024, Tsukui2025}. We use {\tt IMFIT} \citep{Erwin2015} to perform the 2D fittings, incorporating an edge-on disk model and a bulge model. The hybrid PSFs generated in Section~\ref{hybrid} are used. The bulge is modeled with a round Sérsic function \citep{Sersic1968} with $n=4$:
\begin{equation}\label{eq:bulge}
I(R) = I_e \exp\left\{ -b_n \left[ \left( R/R_e \right)^{1/n} - 1 \right] \right\},
\end{equation}
\noindent
where $R_e$ is the half-light radius, $I_e$ is the intensity at $R_e$, and $b_n$ is a function of $n$ determined by the incomplete gamma function. 

We adopt the sech$^2$ vertical profile and the exponential radial profile for the edge-on disk model\footnote{Our measured scale height from Eq.~(\ref{Eq:sech2}) is twice that derived from the default sech$^2$ function in {\tt IMFIT}.}: 
\begin{equation}\label{Eq:sech2}
  I(R,h) = I_0 \cdot (R/R_d) \cdot K_1(R/R_d) \cdot {\rm sech}^{2}(h/h_0),
\end{equation}
\noindent
where $h_0$ is the sech$^2$ scale height. The Sérsic-sech$^2$ decomposition is carried out for each galaxy in all available filters. The galaxy center is a free parameter to fit, as this can mitigate the effects of dust lanes (see Section~\ref{sect:rest}).  The uncertainties of the best-fit parameters are provided by {\tt IMFIT} as the statistical uncertainties obtained from evaluating the structure of the $\chi^2$ minimum. The F115W and F150W images generally have lower $S/N$ than the F277W and F444W images, as the stellar emission of galaxies typically peaks at the NIR.  At high redshift, some galaxies observed in F115W or F150W are indistinguishable from, or only marginally above, the background noise. To ensure robust measurements, we exclude any filter whose mean flux within the galaxy segmentation falls below the background noise level and instead adopt filters with sufficiently high $S/N$.  The scale height derived from the Sérsic-sech$^2$ fitting represents the scale height averaging thin and thick components when present. Figure~\ref{fig:example} presents representative fitting results for galaxies at $z=0.54$ and $z=2.45$.  We further interpolate the measurements to a fixed rest-frame wavelength of 1\,\micron\ to mitigate the effects of dust extinction and spatial variation in stellar populations, as detailed in Section~\ref{sect:rest}. The correction for projection biases arising from deviations from perfect edge-on orientation is described in Section~\ref{sect:incl}.

\begin{figure}
\centering
\includegraphics[width=0.9\columnwidth]{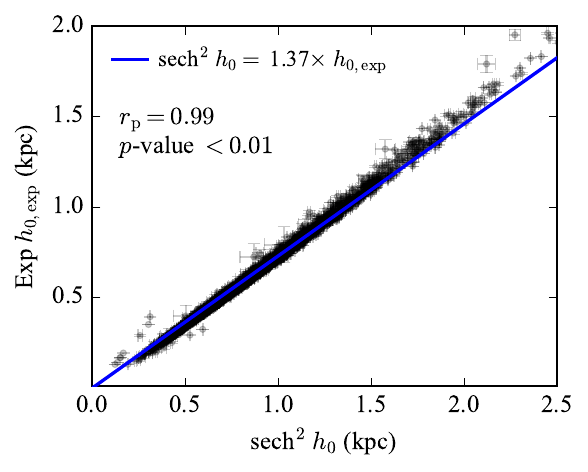}
\caption{Correlation between the scale heights derived from the sech$^2$ function ($h_0$) and the exponential function ($h_{0.\rm exp}$). The two measurements are tightly correlated, with $h_0$ being on average 1.37 times as large as $h_{0,\rm exp}$. }
\label{fig:exp_sech2}
\end{figure}

For comparison, we also measure the exponential scale height ($h_{0,\rm{exp} }$). To minimize degeneracy between components, we fix the bulge parameters to those obtained from the Sérsic-sech$^2$ fit and then refit the image using an exponential vertical profile: 
\begin{equation}\label{Eq:exp}
  I(R,h) = I_0 \cdot (R/R_d) \cdot K_1(R/R_d) \cdot \exp(-h/h_{0,\rm{exp}} ).
\end{equation}
\noindent 
We use $h_{0,\rm{exp} }$ instead of $h_{0,\rm{exp}}/2$ in Eq.~(\ref{Eq:exp_}), as $h_{0,\rm{exp} }$ is more commonly adopted in the literature.
Figure~\ref{fig:exp_sech2} presents the correlation between the two measurements, which follows a tight linear relation described by 
\begin{equation}\label{exp_sech2}
  {\rm sech}^2~h_0 = 1.37\times {\rm exponential}~h_{0,\rm{exp}}.
\end{equation}
\noindent
The Pearson correlation coefficient between sech$^2$ $h_0$ and exponential $h_{0,\rm{exp}}$ is $r_{\rm p}=0.99$, indicating a statistically significant correlation ($p$-value $< 0.01$).  Therefore, our results remain unaffected by whether the vertical profile is described using a sech$^2$ or an exponential function. In the rest of this work, the term ``scale height'' denotes the sech$^2$ scale height unless the exponential scale height is specified.

In addition to the 2D method, one-dimensional (1D) fitting has also been used to measure scale heights \citep{Elmegreen2017, Ranaivoharimina2024, HamiltonCampos2023}. The 1D method is able to obtain scale height as a function of radius.  We evaluate the consistency between $h_0$ derived using 1D and 2D approaches. For the 1D method, we first subtract the best-fit bulge component (if present) and rotate the galaxy so that its major axis aligns with the image $x$-axis. We then extract vertical intensity profiles along the galaxy major axis. Profiles with peak intensities below three times the background noise are excluded. Each vertical profile is fitted with a 1D ${\rm sech}^2$ function, convolved with the line-spread function (LSF), from which we derive the radial profile of scale height. The LSF is constructed by convolving a one-pixel line with a PSF following the strategy in \citet{Elmegreen2017}. The profiles are stacked to create a composite profile. A final ${\rm sech}^2$ fit to the stacked profile yields the average $h_0$ from the 1D method, with weighting similar to that of the 2D fitting.

\begin{figure}
\centering
\includegraphics[width=0.9\columnwidth]{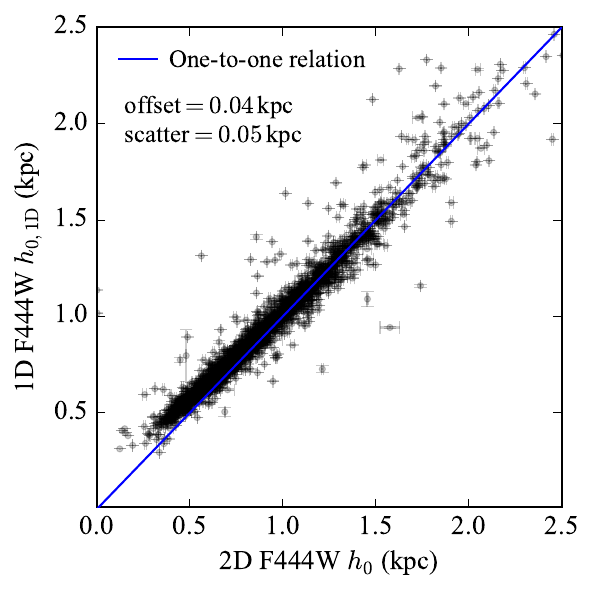}
\caption{Comparison of sech$^2$ scale heights derived from 1D ($h_{0, {\rm 1D}}$) and 2D ($h_0$) fitting methods. The median difference and corresponding standard deviation are displayed at the top.}
\label{fig:2D1D}
\end{figure}

A comparison of the 1D and 2D measurements is shown in Figure~\ref{fig:2D1D}. The median difference between the two is 0.04\,kpc, with a standard deviation of 0.05\,kpc. These small values indicate that the scale heights derived from the 1D and 2D methods are consistent when similar weighting schemes are applied. The radial variation of scale height will be discussed further in Section~\ref{sec:results}.

\begin{figure*}
\centering
\includegraphics[width=2\columnwidth]{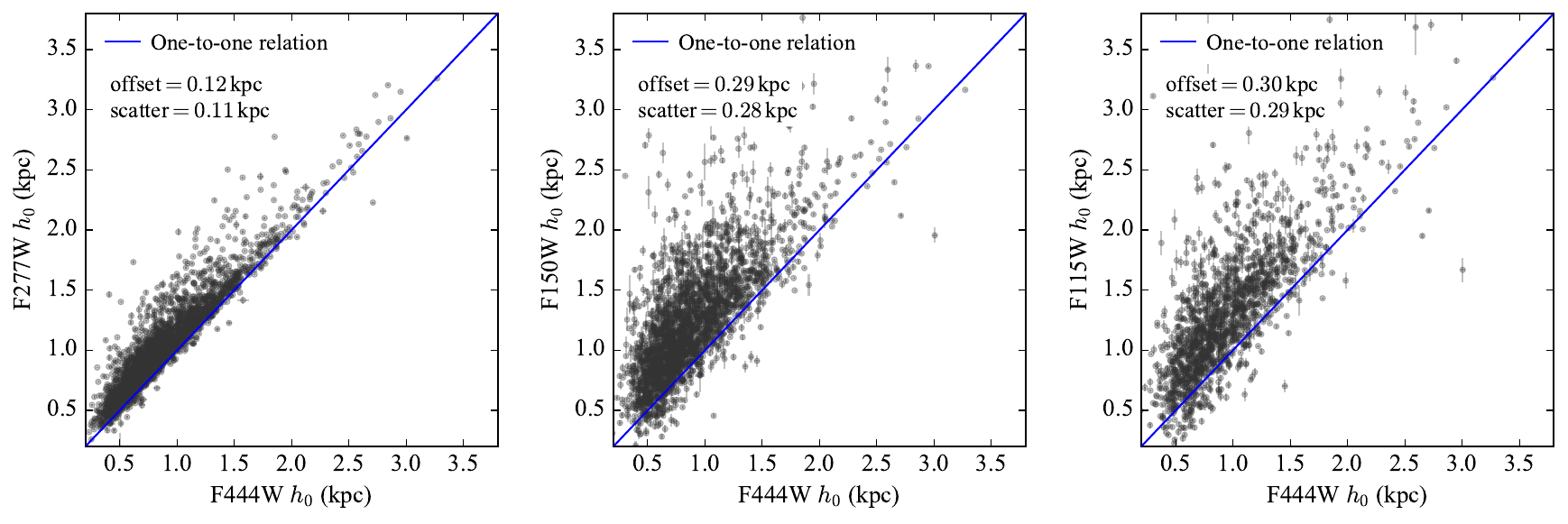}
\caption{Correlation between scale heights measured in F444W and those in F277W, F150W, and F115W. The blue line marks the one-to-one relation. The median offset relative to F444W and scatter are indicated.}
\label{fig:hz_wave}
\end{figure*}

\begin{figure*}
\centering
\includegraphics[width=2\columnwidth]{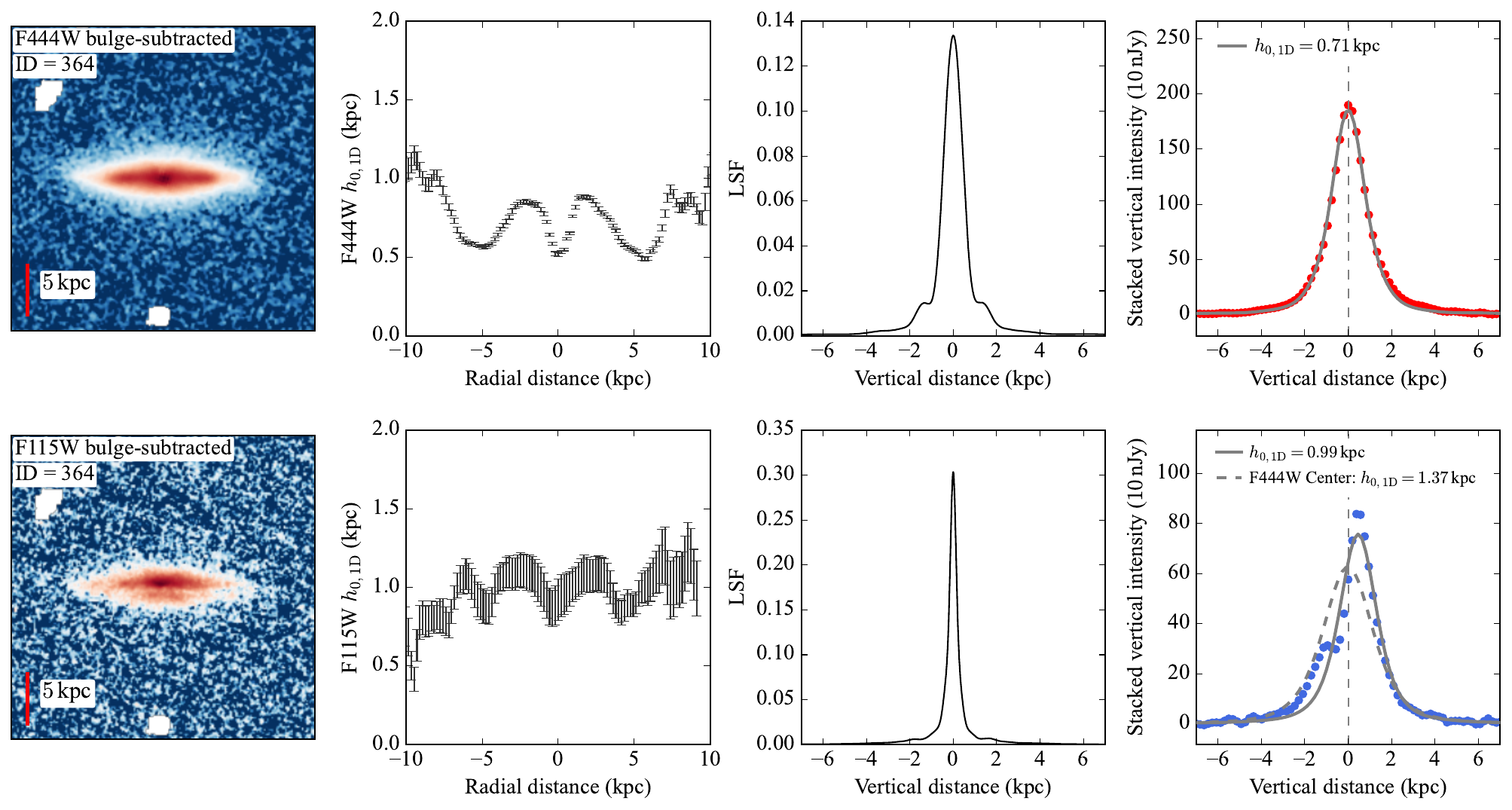}
\caption{Illustration explaining why the scale height derived from the bluer filter appears thicker than that from the redder filter. From left to right: the bulge-subtracted image, the radial profile of scale height, the line-spread function (LSF) constructed from the PSF, and the 1D sech$^2$ fit to the stacked vertical light profile. The top and bottom rows correspond to the F444W and F115W filters, respectively. In the final column, solid curves mark the best fits with all parameters free, and the dashed curve marks the fit with the center fixed to that in F444W.}
\label{fig:blue_thick}
\end{figure*}

\begin{figure}
\centering
\includegraphics[width=1\columnwidth]{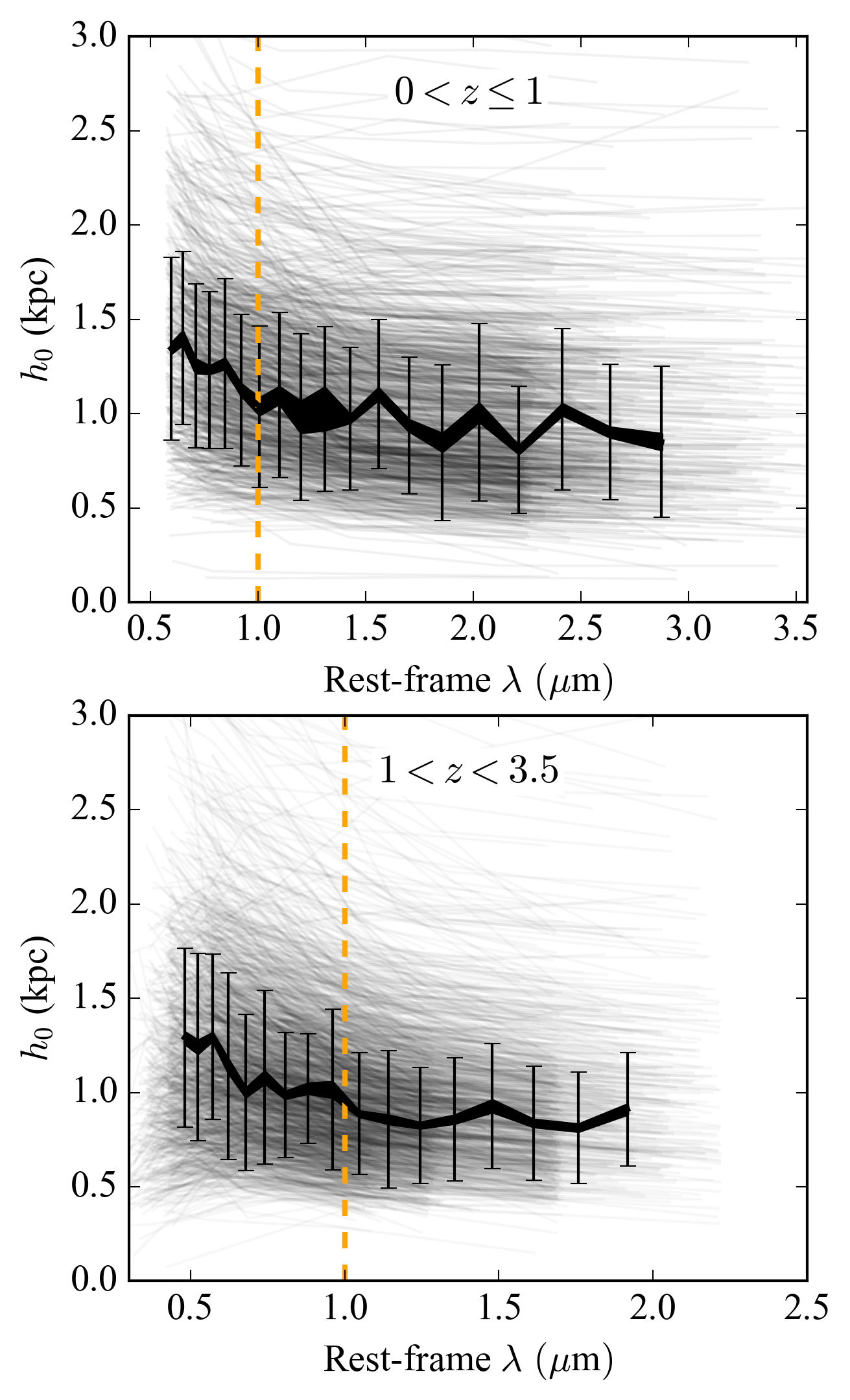}
\caption{Measured scale heights ($h_0$) as a function of rest-frame wavelength. Results for individual galaxies are shown as background gray curves. The data are grouped into 20 bins in rest-frame wavelength. The center and width of the black region (curve) indicate the median values and their uncertainties, while the error bars represent the scatter among individual measurements. The vertical orange dashed line marks the rest-frame 1\,$\micron$. The top and bottom panels show results for $0<z\leq1$ and $1<z<3.5$, respectively. Dust extinction produces a central depression in the vertical light profile, especially at short rest-frame wavelength, which causes the fitted scale height to be overestimated.}
\label{fig:hz_rest}
\end{figure}

\section{Results} \label{sec:results}

\subsection{Wavelength dependence of scale height and the correction}\label{sect:rest}

Since younger stellar populations dominate the thin disks of galaxies, observations at shorter rest-frame wavelengths without dust extinction are expected to reveal thinner disks with smaller scale heights. However, for edge-on systems, particularly the actively star-forming disks in our sample, dust lanes along the mid-plane substantially attenuate the flux especially at shorter wavelengths. This obscuration impacts the apparent vertical light profile and can overestimate the measured scale height \citep{Bizyaev2009}. As described by Eq.~(\ref{Eq:sech2}), the scale height is defined as the height at which the intensity falls to sech$^2$(1) of the central value, rather than as an absolute geometric thickness.  When the mid-plane flux is suppressed by a dust lane, the observed profile appears artificially broadened, leading to an overestimation of the intrinsic scale height.

In Figure~\ref{fig:hz_wave}, we compare the vertical scale heights ($h_0$) measured in the F444W band with those measured in F277W, F150W, and F115W. The scale heights derived from the bluer filters are systematically larger than those from redder filters. Quantitatively, the $h_0$ measured in F277W, F150W, and F115W exceeds those in F444W by $0.12\pm0.11$\,kpc, $0.29\pm0.28$\,kpc, and $0.30\pm0.29$\,kpc, respectively.

To visualize how dust attenuation overestimates the measured scale height, we plot in Figure~\ref{fig:blue_thick} examples of the 1D vertical profile fitting (see Section~\ref{h_measure} for the method) for a galaxy observed in F444W (top) and F115W (bottom).  The first column shows the bulge-subtracted images, where the prominent dust lane in F115W is clearly visible. The second column presents the radial profile of $h_0$, showing systematically larger values in F115W than in F444W, consistent with the 2D fitting results for the same galaxy shown in Figure~\ref{fig:example}. The third column shows the corresponding LSFs, while the final column compares the vertical light profiles. The F444W vertical light profile is symmetric and smooth, whereas the F115W profile exhibits a flux depression caused by the dust lane near the center. This depression shifts the apparent center and broadens the observed distribution, producing an overestimated scale height. When all fitting parameters are allowed to vary freely, we obtain $\h=0.71$\,kpc in F444W and $\h=0.99$\,kpc in F115W. If the galaxy center in F115W is fixed to that of F444W, the fitted value in F115W increases further to $1.37$\,kpc. Setting the center free for the fitting can mitigate the  effect caused by dust lane, which motivates our choice of leaving the center unconstrained in the bulge-edge-on-disk decomposition (Section~\ref{sec:methods}). In addition to dust extinction, spatial variations in the stellar population, such as younger stars concentrated near the mid-plane and older stars located at larger vertical distances, can in principle also affect the scale height measured at different wavelengths, although this effect is less significant than dust extinction.

Figure~\ref{fig:hz_rest} presents the dependence of the measured \h\ on rest-frame wavelength, with results shown separately for $0<z\leq1$ (top) and $1<z<3.5$ (bottom). The median values and their uncertainties reveal a clear trend: scale heights systematically decrease toward longer rest-frame wavelengths, where dust extinction becomes weaker. This dependence on rest-frame wavelength should have also incorporated the effects of spatial variations in the stellar population. The measured $h_0$ at rest-frame 5000\,\AA\ is overestimated by roughly 30--40\%, whereas at rest-frame $1\,\micron$, the impact of extinction is small.

This wavelength dependence therefore acts as a confounding factor when a single filter is used to study the redshift evolution of scale height, if the filter traces rest-frame optical to UV wavelengths at high redshift but NIR wavelengths at low redshift. To isolate the true redshift evolution, we determine $h_0$ for each galaxy at a fixed rest-frame wavelength of $1\,\micron$ ($h_{0,1\micron}$) by fitting a power law to the $h_0$ versus rest-frame wavelength, using the uncertainties from 2D fitting as weights. A similar wavelength dependence affects the disk scale length, $R_d$, owing to the inside-out growth of galaxies \citep{vanderWel2014, Lilly2016}. To ensure consistency, we likewise derive $R_d$ at 1\,$\micron$, $R_{d,1\micron}$.  The effective radius of the disk component at 1\,\micron\ ($R_{e,\rm{disk},1\micron}$) is converted from $R_{d,1\micron}$.

We adopt $1\,\micron$ as the reference wavelength as extinction effects are largely mitigated at this wavelength, and it remains accessible at high redshift with the F444W filter. 
At longer rest-frame wavelengths ($\gtrsim1.5\,\micron$), $h_0$ gradually converges to values $\sim$10\% lower than $h_{0,1\micron}$ at both low and high redshifts. Nevertheless, by anchoring both $h_0$ and $R_d$ to the same rest-frame wavelength, our analysis mitigates wavelength-dependent effects, including those caused by dust extinction and stellar population effects, and enables a clean and uniform comparison of disk structure across redshift. The measurements of scale heights and scale lengths can be accessed at Zenodo \citep{Yu2026Zenodo}.

\begin{figure*}
\centering
\includegraphics[width=2\columnwidth]{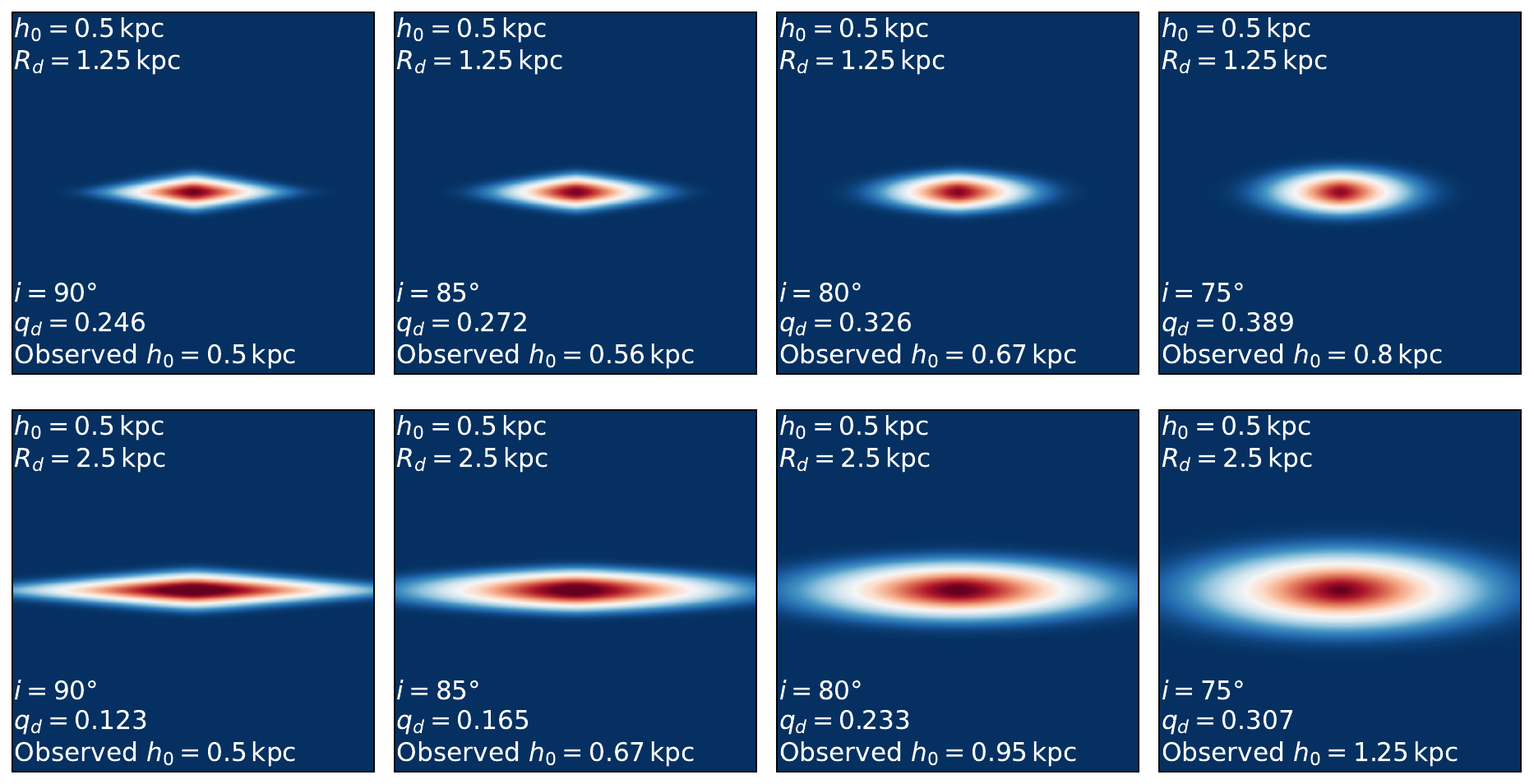}
\caption{Illustration of the projection effect caused by a deviation from a perfect edge-on inclination on scale height $h_0$ measurements. The top and bottom rows present the analysis for models with intrinsic relative scale height $h_0/R_d = 0.4$ and $h_0/R_d = 0.2$, respectively. Columns 1 to 4 correspond to inclination angles of $i = 90\degr$, $85\degr$, $80\degr$, and $75\degr$, respectively. The intrinsic properties are listed at the top in each panel, while the observed properties are listed at the bottom. The observed $h_0$ is more strongly overestimated when the disk deviates more from $i=90$ and has a lower relative scale height. }
\label{fig:impact_incl}
\end{figure*}

\begin{figure*}
\centering
\includegraphics[width=2\columnwidth]{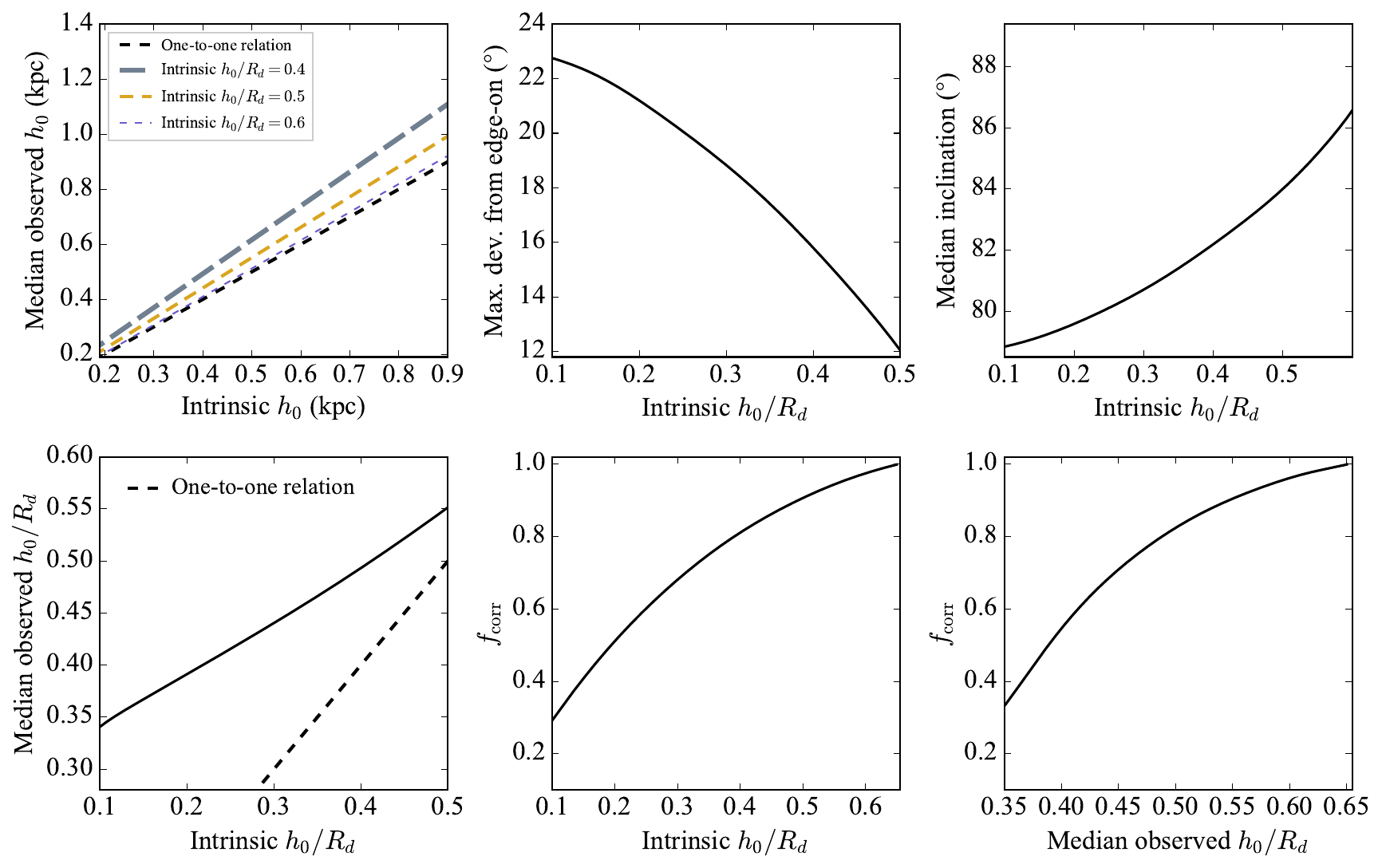}
\caption{Projection effects of the deviation from a perfect edge-on inclination on scale height measurement and the correction curve in a disk galaxy sample with $q_d<0.4$. The top row shows the relation between the median observed and intrinsic $h_0$ (left), and the dependence of the maximum deviation from a perfect edge-on inclination required for a disk to satisfy $q_d < 0.4$ (middle) and the median inclination angle (right) on intrinsic $h_0/R_d$. The bottom row shows the median observed $h_0/R_d$ (left) and the correction factor $f_{\rm corr}$ (middle) as functions of intrinsic $h_0/R_d$, and $f_{\rm corr}$ as a function of median observed $h_0/R_d$ (right).
}
\label{fig:cor_func}
\end{figure*}

\subsection{Projection effects of deviations from a perfect edge-on inclination and the bias correction}\label{sect:incl}

A randomly oriented three-dimensional (3D) galaxy disk is unlikely to be perfectly edge-on. A deviation from perfect edge-on inclination ($i = 90\degr$) amplifies the scale height measured in a 2D projection image, which is done by assuming a perfect edge-on inclination. This constitutes a projection effect overestimating the scale height. Consequently, correcting for this effect is essential to derive the intrinsic scale heights and their intrinsic cosmic evolution, particularly for our sample, which is selected based on a cut in the observed disk axis ratio $q_d<0.4$.

A constant correction factor of 0.78 has been proposed to address this bias \citep{HamiltonCampos2023}. However, this correction is not accurate enough, because the influence of the deviation from a perfect edge-on inclination on scale height measurement actually depends on the relative scale height of the disk, demonstrated below. 
The relative scale height (or relative thickness) is defined as
\begin{equation}
  {\rm  relative~scale~height}=h_0/R_d.
\end{equation}
\noindent
We generate simulated projected 2D images of disks at specific inclination by using {\tt IMFIT} \citep{Erwin2015}, which performs line-of-sight integration of 3D disk models with exponential radial and ${\rm sech}^2$ vertical profiles. For illustration purpose, we consider two disks with $h_0/R_d=0.4$ and $0.2$, respectively, and rotate them at $i=90\degr$, $85\degr$, $80\degr$, and $75\degr$. The projected 2D images with intrinsic and observed values of the disk parameters are shown in Figure~\ref{fig:impact_incl}. For the model with $h_0/R_d=0.4$, when inclination is set from $i=90\degr$ to $i=75\degr$, $h_0$ is amplified from 0.5\,kpc to 0.8\,kpc, and $q_d$ is amplified from $0.246$ to $0.389$. But for $h_0/R_d=0.2$, $h_0$ is amplified from 0.5\,kpc to 1.25\,kpc and $q_d$ is amplified from $0.123$ to $0.307$. We can learn two key points from this test: (1) the scale height of a disk with lower relative scale height (e.g., lower $h_0$ at fixed $R_d$, or larger $R_d$ at fixed $h_0$) suffers more severely from the projection effect; (2) a disk with lower relative scale height is less likely to meet the cut in $q_d$ when selecting a sample.

Next we quantify the impact of this projection effect and derive a new bias correction method. We use {\tt Imfit} to construct a suite of disk models spanning a range of $h_0/R_d$ and inclination angles. For each model, we perform fits to measure $h_0$, $R_d$, and $q_d$. The measurement of $R_d$ remains robust, but $h_0$ is overestimated. To derive the correction for measured $h_0$, we apply our sample selection criterion by excluding models with $q_d \ge 0.4$ and compute the median inclination angle and median observed $h_0$, weighted by the inclination probability distribution $P(i) \propto \sin(i)$.

The results are presented in Figure~\ref{fig:cor_func}. The top-left panel illustrates the median observed value of \h\ of disks with larger intrinsic \hR\ are less overestimated. The top-middle panel demonstrates that disks with larger intrinsic \hR\ ratios can have smaller maximum deviations from $i = 90\degr$ before exceeding the $q_d$ threshold, simply because relative thinner disk has smaller axis ratio. The top-right panel present the median inclination angle as a function of intrinsic \hR. The bottom-left panel show the conversion between intrinsic and median observed \hR. The bottom-middle and bottom-right panel shows the derived statistical correction factor $f_{\rm corr}$, defined as:
\begin{equation}
  f_{\rm corr} = \frac{{\rm median~intrinsic}~h_0}{{\rm median~observed}~h_0},
\end{equation}
\noindent
plotted as a function of intrinsic and median observed \hR, respectively. Disks with larger \hR\ ratios exhibit correction factors closer to 1, because (1) their \h\ are less affected for a given inclination compared to those with lower intrinsic \hR\ (see Figure~\ref{fig:impact_incl}) and (2) their maximum deviation from a perfect edge-on inclination before meeting the cut in $q_d$ is smaller (top-middle panel).  The variation in $f_{\rm corr}$ highlights the importance of the newly developed correction method, as a redshift evolution of $h_0/R_d$ indeed exists, as shown in the next section. Table~\ref{tab} shows the parameters for deriving the correction factors in samples selecting using the axis ratio cut $q_d<0.4$ as in this work.  Table~\ref{tab_03} provides values for samples selecting using $q_d<0.3$. If a smaller critical axis ratio is used, the $f_{\rm corr}$ is closer to unity.

\begin{deluxetable}{ccccc}
\tablewidth{0pt}
\tablecaption{
Parameters for deriving the correction factors for removing the projection effect caused by deviations from edge-on orientation in samples selected with axis ratio cut, $q_d<0.4$. \label{tab}}
\tablehead{
\colhead{Observed $h_0/R_d$} & \colhead{Intrinsic $h_0/R_d$} & \colhead{Max. Deviation $i$ from edge-on} & \colhead{Median $i$} & \colhead{$f_{\rm corr}$}\\
\colhead{} & \colhead{} & \colhead{($\degr$)} & \colhead{($\degr$)} & \colhead{} \\
\colhead{(1)} & \colhead{(2)} & \colhead{(3)} & \colhead{(4)} & \colhead{(5)}
}
\startdata
0.350 & 0.116 & 22.6 & 78.9 & 0.332 \\
0.375 & 0.166 & 21.9 & 79.3 & 0.442 \\
0.400 & 0.218 & 20.8 & 79.8 & 0.546 \\
0.425 & 0.270 & 19.6 & 80.3 & 0.635 \\
0.450 & 0.319 & 18.3 & 81.0 & 0.709 \\
0.475 & 0.366 & 16.9 & 81.7 & 0.771 \\
0.500 & 0.412 & 15.4 & 82.4 & 0.824 \\
0.525 & 0.456 & 13.8 & 83.1 & 0.868 \\
0.550 & 0.498 & 12.2 & 83.9 & 0.905 \\
0.575 & 0.538 & 10.3 & 84.9 & 0.935 \\
0.600 & 0.577 & ~8.3  & 85.9 & 0.962 \\
0.625 & 0.614 & ~5.9  & 87.1 & 0.982 \\
0.650 & 0.649 & ~1.8  & 89.2 & 0.999 \\
\enddata
\tablecomments{Col. (1): Median observed relative scale heights. Col. (2): Intrinsic relative scale heights. Col. (3): Maximum deviation from a perfect edge-on inclination before reaching $q_d\geq 0.4$. Col. (4): Median inclination. Col. (5): Correction factor for the median observed scale height accounting for bias due to deviation from a perfect edge-on orientation.}
\end{deluxetable}

\begin{deluxetable}{ccccc}
\tablewidth{0pt}
\tablecaption{Parameters for deriving the correction factors for removing projection effect caused by deviations from edge-on orientation in samples selected with axis ratio cut, $q_d<0.3$. \label{tab_03}}
\tablehead{
\colhead{Observed $h_0/R_d$} & \colhead{Intrinsic $h_0/R_d$} & \colhead{Max. Deviation $i$ from edge-on} & \colhead{Median $i$} & \colhead{$f_{\rm corr}$}\\
\colhead{} & \colhead{} & \colhead{($\degr$)} & \colhead{($\degr$)} & \colhead{} \\
\colhead{(1)} & \colhead{(2)} & \colhead{(3)} & \colhead{(4)} & \colhead{(5)}
}
\startdata
0.26 & 0.090 & 16.6 & 81.8 & 0.348 \\
0.28 & 0.126 & 16.1 & 82.0 & 0.450 \\
0.30 & 0.166 & 15.3 & 82.4 & 0.553 \\
0.32 & 0.206 & 14.3 & 82.9 & 0.645 \\
0.34 & 0.246 & 13.3 & 83.4 & 0.722 \\
0.36 & 0.283 & 12.1 & 84.0 & 0.787 \\
0.38 & 0.319 & 10.9 & 84.6 & 0.839 \\
0.40 & 0.353 & ~9.7 & 85.2 & 0.883 \\
0.42 & 0.387 & ~8.4 & 85.8 & 0.920 \\
0.44 & 0.418 & ~6.8 & 86.6 & 0.951 \\
0.46 & 0.448 & ~5.1 & 87.5 & 0.975 \\
0.48 & 0.477 & ~2.7 & 88.7 & 0.993 \\
\enddata
\tablecomments{Col. (1): Median observed relative scale heights. Col. (2): Intrinsic relative scale heights. Col. (3): Maximum deviation from a perfect edge-on inclination before reaching $q_d\geq 0.3$. Col. (4): Median inclination. Col. (5): Correction factor for the median observed scale height accounting for bias due to deviation from a perfect edge-on orientation.}
\end{deluxetable}

\begin{figure*}
\centering
\includegraphics[width=2\columnwidth]{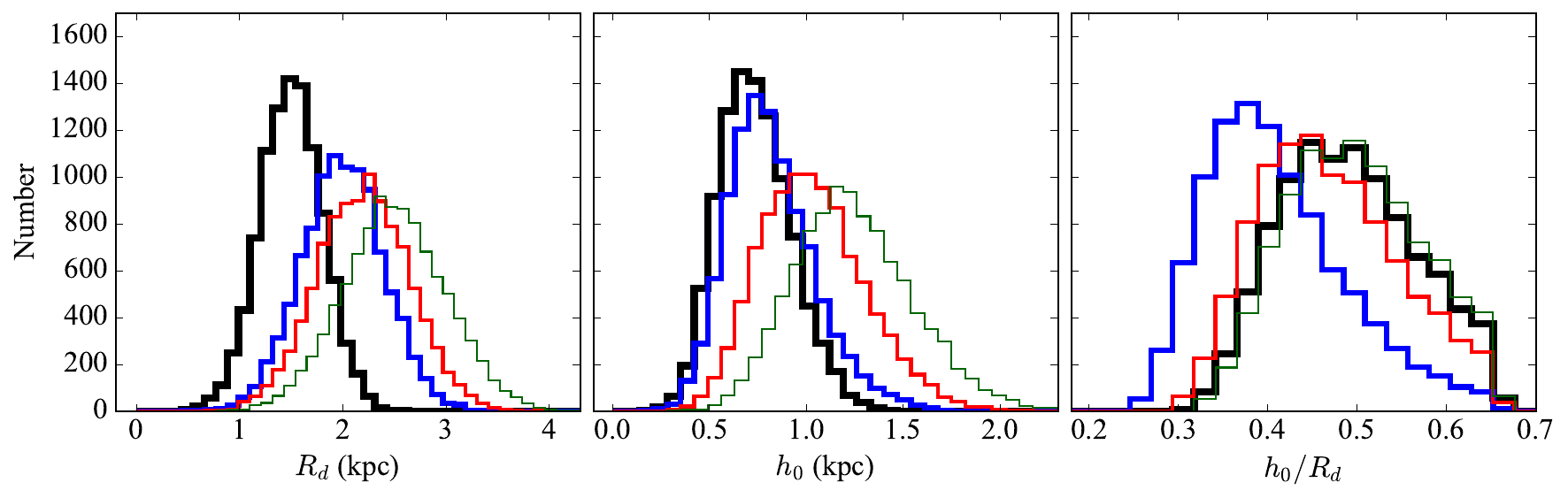}
\caption{Distributions of $R_d$, $h_0$, and $h_0/R_d$ of the four groups of disk models used to verify the effectiveness of our correction method. The four groups are marked in black, blue, green, and red. Each group consists of 10,000 models. A truncation at $h_0/R_d = 0.65$ is applied, as models with values above this threshold exhibit an axis ratio $q_d > 0.4$ even for a perfect edge-on inclination.}
\label{fig:mock}
\end{figure*}

The application of our bias correction is straightforward: we first compute the median observed \hR\ and determine the corresponding $f_{\rm corr}$, and then obtain the corrected median \h\ by multiplying the observed value by $f_{\rm corr}$.  As described above, the correction curve is derived for a given intrinsic $h_0/R_d$. In real observations, however, disks exhibit a distribution of $h_0/R_d$ rather than a single value. To assess the effectiveness of our method under real observational conditions, we create four groups of models, each containing 10,000 models with random inclinations. The distributions of $R_d$, $h_0$, and \hR\ for these four groups are shown in Figure~\ref{fig:mock}. A truncation is observed at $\hR=0.65$ in the \hR\ distribution, as models with \hR\ above this threshold exhibit an axis ratio $q_d \geq 0.4$ even in a perfectly edge-on inclination. For each inclined model, the observed $h_0$ and \hR\ are measured using 2D ${\rm sech}^2$ fitting. For each group, we calculate the median observed $h_0$ and \hR, and apply our new correction method to obtained the bias-corrected \h.

\begin{figure}
\centering
\includegraphics[width=1\columnwidth]{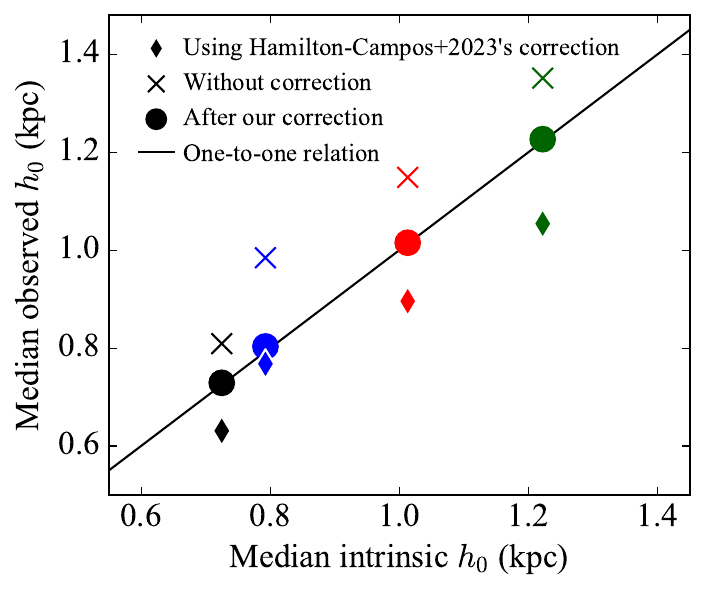}
\caption{Comparison of the median measured scale height ($h_0$), with and without correction, against the intrinsic values. Symbol colors correspond to the models shown in Fig.~\ref{fig:mock}. Crosses represent results without correction, diamonds indicate results using the correction from \citet{HamiltonCampos2023}, and points show results from our new correction method. }
\label{fig:demo}
\end{figure}

Figure~\ref{fig:demo} presents a comparison of the median measured $h_0$, with and without correction, against the intrinsic values. The results without correction are shown as crosses. As expected, the observed $h_0$ values are systematically overestimated compared to their intrinsic values when no correction is applied. Furthermore, the degree of overestimation varying across different model groups, suggesting there is no constant factor that can entirely correct for the effects.  If a constant correction factor of 0.78, as suggested by \citet{HamiltonCampos2023}, is applied, the corrected $h_0$ values (shown as diamonds) do not reproduce the intrinsic values.  In contrast, our correction method, represented by solid points, yields $h_0$ values that align closely with the one-to-one relation, demonstrating the effectiveness of our approach in accounting for the projection effect caused by deviations from a perfect edge-on inclination. We emphasize that our method is statistical in nature and can be applied only to a sample, not to an individual source.

\begin{figure}
\centering
\includegraphics[width=1\columnwidth]{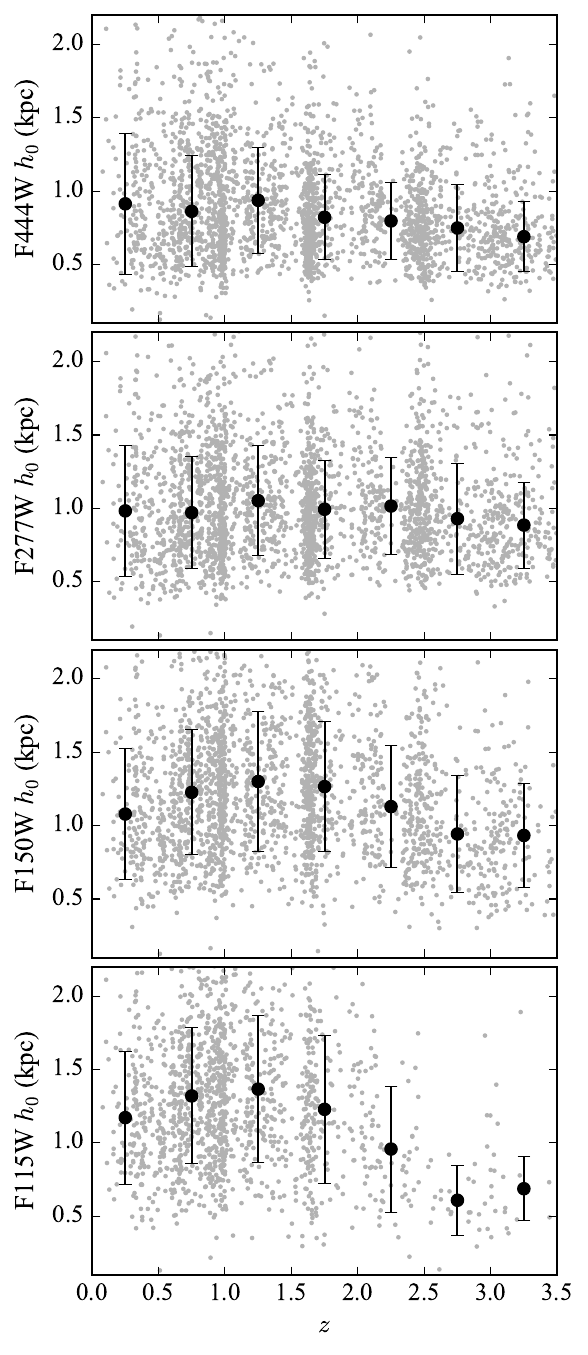}
\caption{Scale heights derived from the F444W, F277W, F150W, and F115W bands as a function of redshift. The black points and associated error bars mark the median values and the scatter among individual measurements. At high redshift, the number of galaxies included in the F115W and F150W measurements decreases because in these filters the galaxies become too faint in surface brightness to yield reliable measurements (see Section~\ref{sec:methods}). }
\label{fig:hz_z_wave}
\end{figure}

\begin{figure*}
\centering
\includegraphics[width=2\columnwidth]{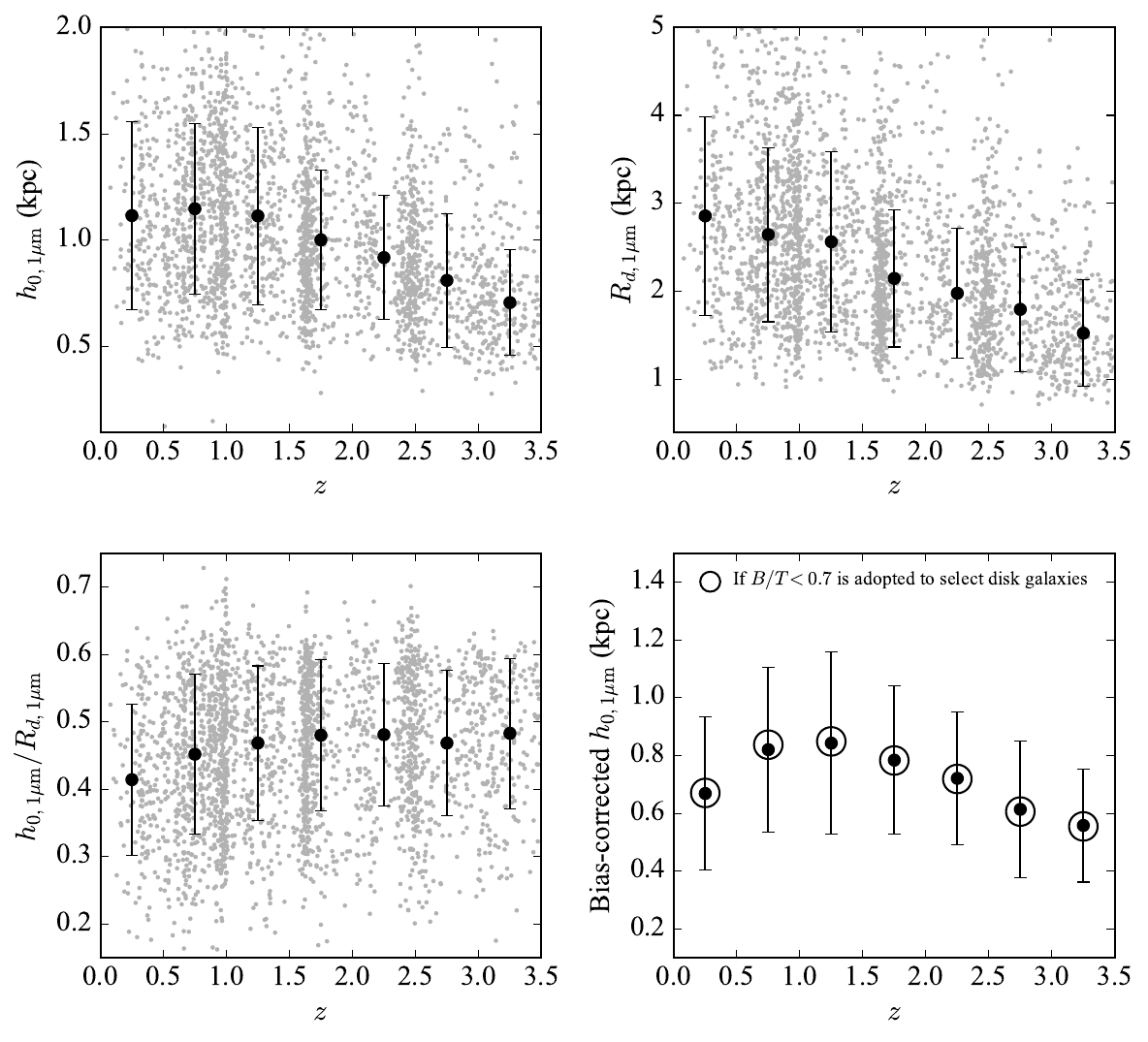}
\caption{Measured at a fixed rest-frame wavelength of 1,\micron, the scale height ($h_{0,1\micron}$), scale length ($R_{d,1\micron}$), relative scale height ($h_{0,1\micron}/R_{d,1\micron}$), and the bias-corrected $h_{0,1\micron}$ (accounting for projection effects) are plotted as functions of redshift. Black points with error bars indicate the median values and the scatter among individual measurements across 7 redshift bins.}
\label{fig:hz_L_z}
\end{figure*}

\begin{figure}
\centering
\includegraphics[width=0.95\columnwidth]{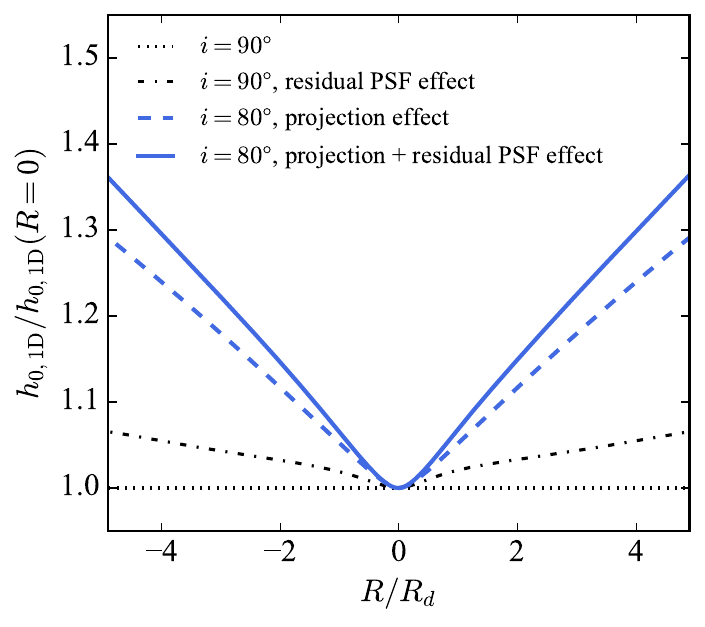}
\caption{Artificial increases of scale height as a function of normalized radius ($R/R_d$). The increase is primarily driven by deviations from a perfectly edge-on orientation, with a minor contribution from residual PSF effects. The black dotted and dot-dashed curves represent the results for a perfectly edge-on disk ($i=90\degr$) without and with residual PSF effects, respectively, while the blue dashed and solid curves show the corresponding results for a disk inclined at $i=80\degr$.
}
\label{fig:radial_hz}
\end{figure}

\begin{figure}
\centering
\includegraphics[width=0.95\columnwidth]{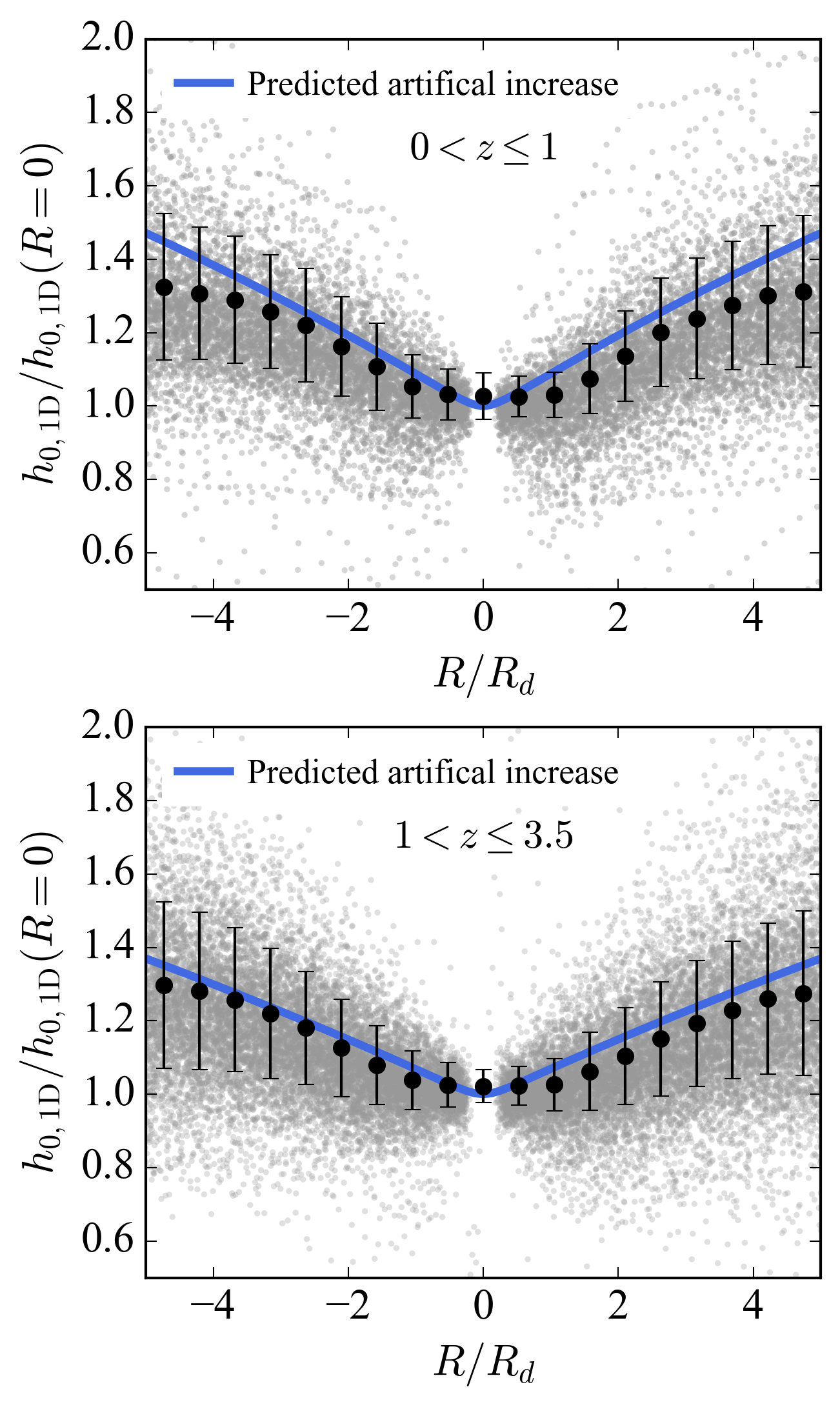}
\caption{Normalized measured scale height profiles for galaxies at $0<z\leq1$ (top) and $1<z\leq3.5$ (bottom). Black points with error bars show the median and scatter. The solid blue curves indicate the predicted artificial increase due to median residual PSF effects and the median deviation from a perfectly edge-on inclination (predicted median $i=80\fdg9$ and $81\fdg7$, respectively). Note that $5R_d\approx3R_e$.
}
\label{fig:rad_grad}
\end{figure}

\begin{figure*}
\centering
\includegraphics[width=2\columnwidth]{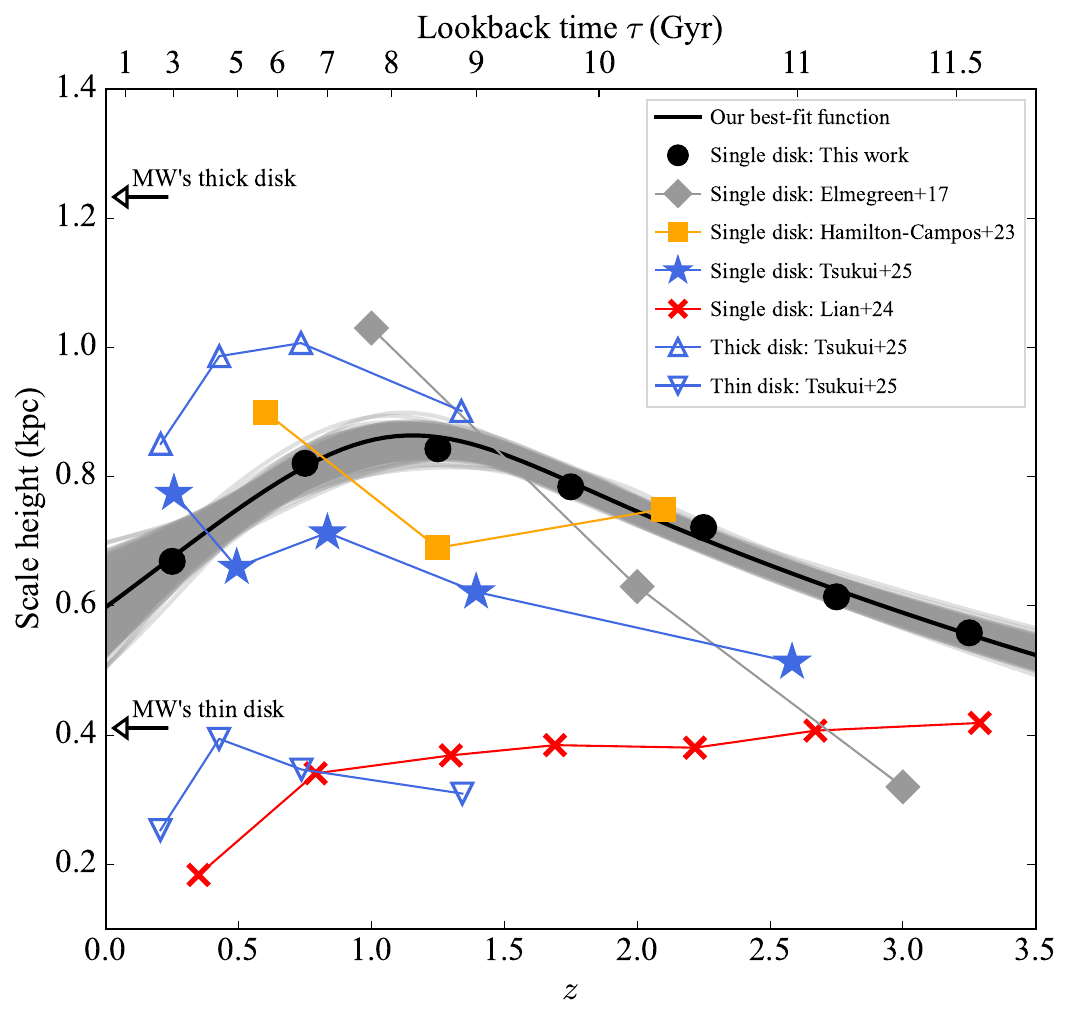}
\caption{Redshift evolution of our bias-corrected scale heights and comparison with previous studies. Error bars are omitted for clarity. The global disk scale heights are from this work (points; median stellar mass $M_*\approx10^{10.4}\,M_\odot$), \cite{Elmegreen2017} (diamonds; $10^{10}\,M_\odot$), \cite{HamiltonCampos2023} (squares; $10^{9.3}\,M_\odot$), \cite{Tsukui2025} (stars; $10^{9.2}\,M_\odot$), and \cite{Lian2024} (crosses; $10^{9}\,M_\odot$). Results for thick and thin disks are taken from \cite{Tsukui2025} (open upward and downward triangles, respectively; $10^{9.2}\,M_\odot$). The black curve shows the best-fit function to our measurements, while the gray curves, indicating the fitting uncertainty, are the 1000 best-fit functions from 1000 resamplings of the data. The arrows indicate the sech$^2$ scale heights (1.23 and 0.41\,kpc) of the Milky Way’s thick and thin disks, derived by converting the exponential scale heights (0.9 and 0.3\,kpc) reported by \cite{Juric2008} using the conversion in Eq.~(\ref{exp_sech2}).
}
\label{fig:hz_comp_z}
\end{figure*}

\subsection{Redshift evolution of bias-corrected scale heights}\label{sec:bias_remove}

Before examining the bias-corrected scale heights, we first probe the uncorrected measurements derived from the F444W, F277W, F150W, and F115W filters as a function of redshift (Figure~\ref{fig:hz_z_wave}). Only a weak trend is seen in the F444W filter, where the scale height increases slightly toward lower redshift and then remains roughly constant. At high redshift, the number of galaxies included in the F115W and F150W measurements decreases because in these filter the galaxies become too faint in surface brightness to yield reliable measurements and are therefore excluded (see Section~\ref{sec:methods}).   

To remove the bias introduced by variations in rest-frame wavelength, we have computed the scale height ($h_{0,1\micron}$), scale length ($R_{d,1\micron}$), and relative scale height ($h_{0,1\micron}/R_{d,1\micron}$) at a fixed rest-frame wavelength of 1\,$\micron$ (see Section~\ref{sect:rest}). Their redshift dependence is shown in Figure~\ref{fig:hz_L_z}.  The relation between $h_{0,1\micron}$ and redshift is more clearly defined than that derived from single-band measurements, as biases caused by variations in rest-frame wavelength, such as dust extinction and stellar population effects, have been removed. The $R_{d,1\micron}$ exhibits a general increase toward lower redshifts, consistent with the well-established growth of galaxy sizes over cosmic time \citep{vanderWel2014, Allen2025, Yang2025}. The observed $h_{0,1\micron}/R_{d,1\micron}$ decreases with lower redshift, resulting in smaller values of $f_{\rm corr}$ at low redshift. This indicates that the measured scale heights is more overestimated at lower redshifts than at higher redshifts.

To remove the bias introduced by the projection effect, we divide the sample into 7 redshift bins and compute the median observed value of $h_{0,1\micron}/R_{d,1\micron}$ in each bin. These values are used as input to determine the $f_{\rm corr}$ (see Figure~\ref{fig:cor_func} or Table~\ref{tab}), which ranges from 0.6 to 0.8. $f_{\rm corr}$ is then applied to $h_{0,1\micron}$ to obtain the bias-corrected scale height. The bias-corrected $h_{0,1\micron}$ as a function of redshift is presented in Figure~\ref{fig:hz_L_z} (bottom-right) and are listed in Table~\ref{tab:results}.  Therefore, after correcting for the effects of variation in rest-frame wavelength and the projection effect, the median and scatter of the scale heights are $0.56\pm0.03$, $0.61\pm0.04$, $0.72\pm0.03$, $0.78\pm0.03$, $0.84\pm0.04$, $0.82\pm0.03$, and $0.67\pm0.06$\,kpc at $z=3.25$, $2.75$, $2.25$, $1.75$, $1.25$, $0.75$, and $0.25$, respectively. The scatters among individual measurements are 0.2--0.3\,kpc. The overall trend shows an increase in scale height from $z\sim3$ to $z\sim1$, followed by a decline toward lower redshift.  The bias-corrected ratios of $h_{\rm vert, 1\micron}$ to $R_{d,1\micron}$ and their inverse ratios are also presented in Table~\ref{tab:results}.

Furthermore, to examine whether, during sample selection, the adopted cut in $B/T$ used to remove potential elliptical galaxies, which may fall within the star-forming region of the color-color diagram due to measurement uncertainties or are intrinsically star-forming, affects our results, we plot the redshift evolution of the bias-corrected $h_{0,1\micron}$ using a higher $B/T$ threshold of 0.7, shown as the open circles in Figure~\ref{fig:hz_L_z}. The result does not change significantly, suggesting that our conclusions are insensitive to adopting a higher $B/T$ threshold.


\subsection{Artificial radial increase of measured scale heights}\label{artificial_grad}

The projection effect of a deviation from a perfectly edge-on inclination not only causes a systematic overestimation of the global vertical scale height but also produces an artificial radial increase of the measured height toward larger radii, as previously noted by \citet{Bizyaev2014}. In addition, the LSF cannot fully account for the light from the bright central region scattered into the outskirts of the disk, leaving a residual PSF effect that introduces further artificial flaring. Figure~\ref{fig:radial_hz} illustrates these effects using 3D model disks with an intrinsic constant scale height. The measured radial profiles are obtained by fitting 1D vertical light profiles, convolved with a LSF if need, at each radius in the 2D projected image. For a perfectly edge-on disk without PSF convolution, the measured profile normalized by central scale height remains flat at 1. When a PSF is convolved, however, the measured scale height becomes artificially larger even though a LSF is applied when doing the 1D fitting (see Section~\ref{sec:methods}). 
For a disk at $i=70\degr$ without PSF convolution, the measured scale height artificially increases with radius due to projection effects; when the PSF is convolved, this artificial increase becomes slightly stronger owing to the combined impact of projection and residual PSF effects.

To examine whether the disk galaxies in COSMOS-Web exhibit genuine or artificial radial gradients, we measured the vertical scale height at each radius using the 1D method described in Section~\ref{sec:methods}. We refrain from interpolating to rest-frame 1\,$\micron$ because the 1D profiles at individual radii, extracted from 2D flux maps, have relatively lower $S/N$ than global measurements, which would propagate larger uncertainties through interpolation. Instead, we adopted results from the filter probing rest-frame wavelengths just above 1\,\micron, where the effects of dust extinction are small. The galaxies were grouped into two redshift bins: $0 < z \leq 1$ and $1 < z \leq 3.5$. For each galaxy, the radial profile of the scale height was normalized by its central value within one FWHM of the PSF, and the radius was normalized by the disk scale length $R_d$. The resulting normalized profiles are shown in Figure~\ref{fig:rad_grad}, where large points and error bars indicate the median and scatter, respectively. In both redshift ranges, the measured scale height systematically increases with radius, giving the appearance of disk flaring.

However, as shown above, such a trend can arise purely from the projection effect and residual PSF effect. To understand whether this observed increase is physical or artificial, we estimated the median observed $h_{0,1\micron}/R_{d,1\micron}$ for each redshift bin and derived the corresponding intrinsic $h_{0,1\micron}/R_{d,1\micron}$ and median inclination. Using these parameters, we generated PSF-convolved 2D projected models from 3D model disks with constant intrinsic scale height. The 1D fitting method yields the predicted artificial radial trends shown as blue solid curves in Figure~\ref{fig:rad_grad}. The predicted trends closely reproduce the observed radial profiles at both low and high redshift. This agreement indicates that the apparent flaring is primarily a projection artifact, with a minor contribution from residual PSF effects, rather than a genuine structural feature. The implications of this result for the origin of disk scale heights are discussed in Section~\ref{sec:implication}.

\subsection{Comparison with scale height measurements in previous works}\label{sec:comp}

There are several studies on the redshift evolution of disk scale height based on HST or JWST images \citep{ElmegreenElmegreen2006, Elmegreen2017, HamiltonCampos2023, Tsukui2025}. We compare our results with their measurements in Figure~\ref{fig:hz_comp_z}. Our measurements are plotted as black points, with the black curve showing the best-fit relation and the gray curves indicating its uncertainties. The fitting method is described later in Section~\ref{sec:implication}. 

At the Solar Circle of the Milky Way, \citet{Juric2008} reported exponential scale heights of 0.9\,kpc and 0.3\,kpc for the thick and thin disks, respectively, with uncertainties of about 20\%. If these values are multiplied by the commonly used, although inappropriate, factor of 2 to convert to a sech$^2$ scale height, one obtains 1.8\,kpc and 0.6\,kpc. Under this assumption, the Milky Way’s thick disk would appear substantially thicker than the extragalactic values with median less than $\sim\,$1\,kpc, whereas its thin disk would be comparable to the global scale height at $z=0$, suggesting that the Milky Way might be an unusually thick system. However, as shown in Eq.~(\ref{exp_sech2}), the appropriate conversion factor is 1.37 rather than 2. Using this conversion, the sech$^2$ scale heights of the Milky Way’s thick and thin disks become 1.23\,kpc and 0.41\,kpc, respectively, as indicated by arrows in Figure~\ref{fig:hz_comp_z}. These values are a few hundred parsecs higher and lower, respectively, than the global scale heights at $z\approx0$ in this work and in most of the previous studies. This is expected, because the global measurement represents a flux-weighted average of the thin and thick components.

Based on HST/ACS imaging, \citet{Elmegreen2017} measured the scale heights of 107 visually selected edge-on disks with an average stellar mass of $10^{9.5\textendash10.5}\,M_\odot$ at $0.5\leq z\leq 3.5$. They reported a mean scale height of $0.63$\,kpc with a scatter of 0.24\,kpc in the F814W band, which probes rest-frame wavelengths of 0.20--0.54\,\micron\ and is therefore potentially affected by dust extinction. We regrouped their sample into three redshift bins and plot the results as light-gray diamonds in Figure \ref{fig:hz_comp_z}. The data reveal an  increase in scale height from $\sim0.32$\,kpc at $z\approx3$ to $\sim1.03$\,kpc at $z\approx1$.

\cite{HamiltonCampos2023} measured the vertical scale heights of 491 disk galaxies with stellar masses of $M_* = 10^{9\textendash11}\,M_\odot$ (the majority below $10^{10}\,M_\odot$) over the redshift range $0.4 \leq z \leq 2.5$ using HST/ACS F850LP, HST/WFC3 F125W, and F160W imaging. The images are tracing rest-frame optical wavelength range (0.46--0.66\,\micron).  The galaxies were selected with an axis ratio cut of $q < 0.4$, and the scale heights were derived from 1D vertical profile fitting.  A constant correction factor of 0.78 was applied to account for the projection effect, yielding a median scale height of $0.74$\,kpc with a scatter of 0.35\,kpc. The median values of three redshift bins are shown as gray squares in Figure~\ref{fig:hz_comp_z}. They found little to no redshift evolution in the scale height. Nevertheless, the relatively shallow HST observations and their lower spatial resolution may introduce larger measurement uncertainties.

Using F115W imaging, which traces the rest-frame NIR at $z\sim0$ but shifts to the rest-frame UV at $z>3$, \citet{Lian2024}
measured $\operatorname{sech}^2$ scale heights for 191 disk galaxies with a median stellar mass of $\sim10^9\,M_\odot$ over $0.2\le z\le5$, reporting a global median of $0.38$\,kpc with a scatter of 0.13\,kpc. The red crosses in Figure~\ref{fig:hz_comp_z} show their median values in each redshift bin (plotted only for $z<3.5$). Their measured global scale height exhibits only mild evolution at $z>1.5$ but decreases toward lower redshift at $z<1.5$, reaching a surprisingly small value of $\sim0.18$\,kpc at $z=0.35$. Perhaps due to uncertainties caused by small sample statistics, this value is roughly half of the global scale heights reported by \citet{Tsukui2025} for JWST galaxies of similar stellar mass at similar redshift.
\citet{Lian2024} further argued that scale-height measurements are consistent across NIRCam filters from F444W to F115W. However, as demonstrated in Section~\ref{sect:rest} and in \citet{Bizyaev2009}, measurements at shorter rest-frame wavelengths are systematically biased toward larger values due to dust extinction across the mid-plane.

\citet{Tsukui2025} studied 111 nearly edge-on disk galaxies at $z \leq 3$ using F277W, F356W, and F444W images, tracing rest-frame wavelengths of $1$–$2\,\micron$, where dust extinction effect is small. They identified 67 galaxies better fitted with a single-disk component and 44 galaxies requiring two disk components (thin and thick disks), with the earliest two-disk system found at $z = 1.96$. After correcting for mass dependence, the median scale heights at $M_*=10^{9.2}\,M_\odot$ for single disks, thin disks, and thick disks are shown in Figure~\ref{fig:hz_comp_z}.  Consistent with our results, the scale height of single-disk galaxies increases toward lower redshift, with a median value of about $0.66\pm0.07\,$kpc, slightly smaller than our measurements due to their lower stellar masses. In contrast, the thin and thick components of two-disk galaxies show nearly constant, or perhaps slightly decreasing, scale heights with decreasing redshift, with median values of $0.33\pm0.04$ and $0.94\pm0.07\,$kpc, respectively.

\begin{deluxetable*}{ccccccccc}
\tablewidth{0pt}
\tablecaption{Measured scale heights in each redshift bin \label{tab:results}}
\tablehead{
\colhead{} & \colhead{} & \colhead{   } & \colhead{ } & \colhead{Bias-corrected } & \colhead{Bias-corrected } & \colhead{Bias-corrected }\\
\colhead{$z$} & \colhead{$h_{0,1\micron}$} & \colhead{ $h_{0,1\micron}/R_d$  } & \colhead{$f_{\rm corr}$} & \colhead{ $h_{0,1\micron}$} & \colhead{ $h_{0,1\micron}/R_d$} & \colhead{ $R_d/h_{0,1\micron}$}\\
\colhead{} & \colhead{kpc} & \colhead{ } & \colhead{} & \colhead{kpc} & \colhead{} & \colhead{}\\
\colhead{(1)} & \colhead{(2)} & \colhead{(3)} & \colhead{(4)} & \colhead{(5)} & \colhead{(6)} & \colhead{(7)}
}
\startdata
0.25 & $1.12 \pm 0.05~(0.44)$ & $0.41 \pm 0.02~(0.11)$ & $0.599\pm 0.043$ & $0.67 \pm 0.06~(0.27)$ & $0.25 \pm 0.03~(0.07)$ & $4.0 \pm 0.4~(1.1)$\\
0.75 & $1.15 \pm 0.02~(0.40)$ & $0.45 \pm 0.01~(0.12)$ & $0.715\pm 0.018  $ & $0.82 \pm 0.03~(0.29)$ & $0.32 \pm 0.02~(0.09)$ & $3.1 \pm 0.2~(0.9)$\\
1.25 & $1.11 \pm 0.03~(0.42)$ & $0.47 \pm 0.01 (0.11)$ & $0.756\pm 0.025  $ & $0.84 \pm 0.04~(0.31)$ & $0.35 \pm 0.02~(0.09)$ & $2.8 \pm 0.2~(0.8)$\\
1.75 & $1.00 \pm 0.02~(0.33)$ & $0.48 \pm 0.01 (0.11)$ & $0.783\pm 0.014  $ & $0.78 \pm 0.03~(0.26)$ & $0.38 \pm 0.02~(0.09)$ & $2.7 \pm 0.1~(0.7)$\\
2.25 & $0.92 \pm 0.02~(0.29)$ & $0.48 \pm 0.01 (0.10)$ & $0.785\pm 0.016  $ & $0.72 \pm 0.03~(0.23)$ & $0.38 \pm 0.02~(0.08)$ & $2.7 \pm 0.1~(0.6)$\\
2.75 & $0.81 \pm 0.03~(0.31)$ & $0.47 \pm 0.02 (0.11)$ & $0.756\pm 0.032  $ & $0.61 \pm 0.04~(0.24)$ & $0.35 \pm 0.03~(0.08)$ & $2.8 \pm 0.2~(0.7)$\\
3.25 & $0.71 \pm 0.02~(0.25)$ & $0.48 \pm 0.02 (0.11)$ & $0.789\pm 0.026  $ & $0.56 \pm 0.03~(0.20)$ & $0.38 \pm 0.03~(0.09)$ & $2.6 \pm 0.2~(0.7)$\\
\enddata
\tablecomments{Col. (1): redshift; Col. (2)--(3): median, error of the median, and scatter among individual measurements for the measured $h_{0,1\micron}$ and $h_{0,1\micron}/R_d$, respectively; Scatter is shown in the bracket. Col. (4): factor correcting for bias caused by the projection effect; Col. (5)--(7): median, error of the median, and scatter of the distribution for the bias-corrected $h_{0,1\micron}$, $h_{0,1\micron}/R_d$, and $R_d/h_{0,1\micron}$, respectively. 
}
\end{deluxetable*}

\begin{figure}
\centering
\includegraphics[width=0.95\columnwidth]{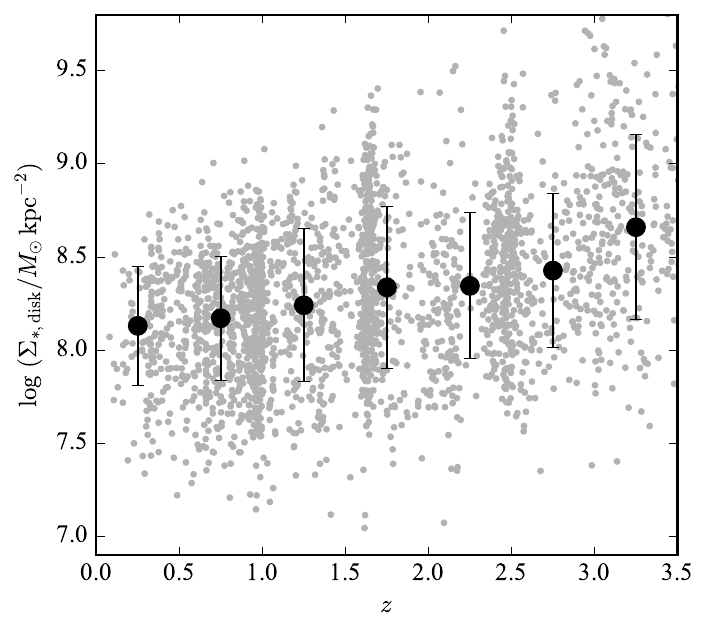}
\caption{Stellar mass surface density of the disk component as a function of redshift. Gray dots represent individual galaxies, while black points with error bars show the median values and the scatter among individual measurements in each redshift bin.
}
\label{fig:mu_mass}
\end{figure}

\begin{figure}
\centering
\includegraphics[width=0.95\columnwidth]{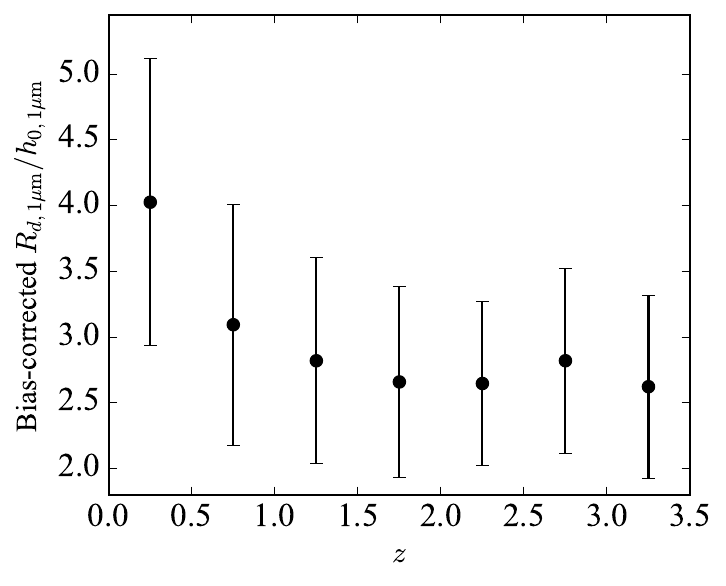}
\caption{Bias-corrected ratio of disk scale length to scale height (the inverse of the relative scale height) as a function of redshift. Black points indicate the median values, and error bars represent the scatter among individual galaxies. The errors on the median are given in Table~\ref{tab:results}.
}
\label{fig:LH}
\end{figure}

\begin{figure*}
\centering
\includegraphics[width=1.95\columnwidth]{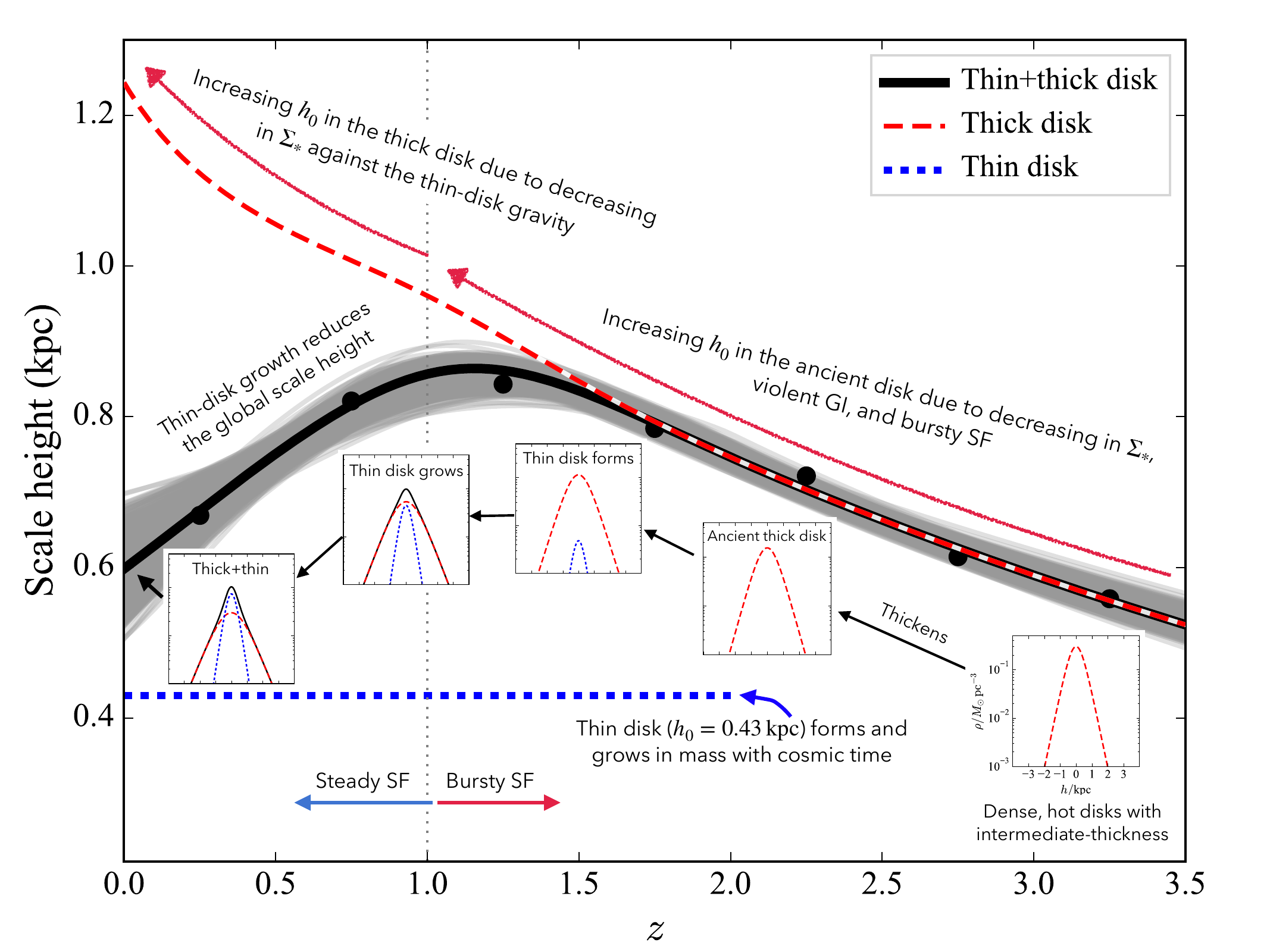}
\caption{Simplified model for the redshift evolution of global disk scale height, averaging the thin and thick disk, at a fixed stellar mass of $M_*=2.5\times10^{10}\,M_\odot$. The solid curve shows the best-fit relation, and the gray curves (shaded region) represent 1000 realizations from resampled datasets. The red dashed curve indicates the best-fit evolution of the thick-disk scale height, while the thin-disk scale height is fixed at 0.43\,kpc. The inset panels indicate how the vertical light profiles of the single-disk-component galaxies, and subsequently thin, thick, and combined disks evolve with redshift. 
}
\label{fig:evo}
\end{figure*}

\begin{figure}
\centering
\includegraphics[width=0.95\columnwidth]{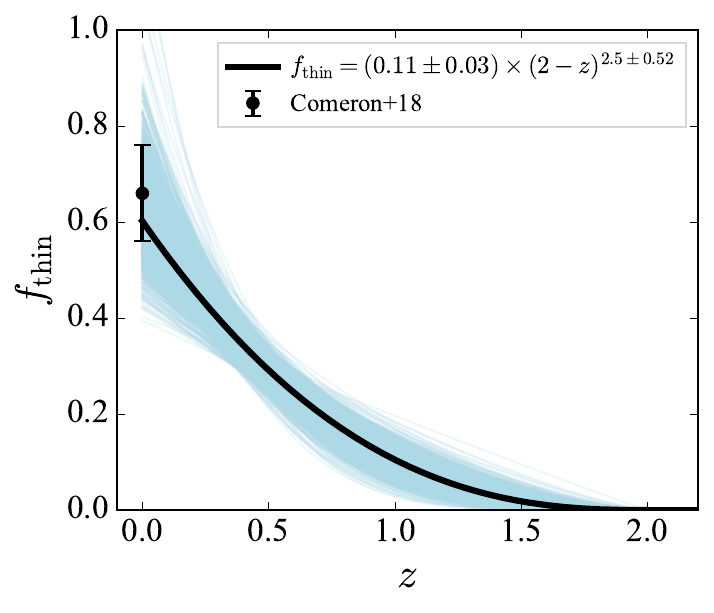}
\caption{Best-fit mass fraction of thin disk as a function of redshift. The solid black curve marks the best-fit function, while the shadowed light blue curves, indicating the fitting uncertainty, are the 1000 fits of the 1000 resamplings of the data. The point and its error bar mark the results from \cite{Comeron2018} for nearby galaxies with stellar mass of $10^{10}$--$10^{10.8}\,M_\odot$ in their sample.
}
\label{fig:f_thin}
\end{figure}

\section{Implications for Galaxy Evolution} \label{sec:implication}

It is important to emphasize that the redshift evolution of disk scale height reported in this work refers to galaxies with a fixed stellar mass of approximately $2.5\times10^{10}\,M_{\odot}$. In reality, galaxies increase their stellar mass over cosmic time; therefore, our analysis does not trace the exact progenitor–descendant relationship of individual systems across redshift. Nevertheless, the results offer valuable clues to the overall cosmic evolution of disk galaxies.

The formation mechanism of thin disks is relatively well understood. Thin disks are composed of stars born from a dynamically cold, rotationally supported gaseous disk. The newborn stars inherit the kinematics of the gas, exhibiting low ratios of velocity dispersion to rotation velocity, and consequently form geometrically thin stellar structures \citep{Forbes2012, YuSijie2023}. However, open questions remain regarding when thin disks first emerged in the Universe and how they evolved over time.

In contrast, the dominant formation mechanism of thick disks remains debated despite decades of research. This uncertainty partly stems from observational limitations, particularly the difficulty in resolving the vertical structure of distant galaxies prior to the JWST era. Broadly speaking, proposed thick-disk formation mechanisms fall into two categories:

(1) Thick disks are born thick. In this picture, stars form with large scale heights of $\sim$1\,kpc either at birth or shortly thereafter. This may occur through rapid stellar scattering driven by violent gravitational instabilities in gas-rich, clumpy high-redshift disks \citep{Bournaud2009, Silva2020, Silva2021}, combined with the dissolution of massive clumps \citep{Elmegreen2006, Elmegreen2017}. It may also arise from star formation within turbulent high-redshift gaseous disks \citep{Romeo2010, Romeo2014}, where giant molecular clouds (GMCs) have sizes comparable to clumps, typically $\sim$300\,pc \citep{Elmegreen2013, DZ2017, Cava2018, Claeyssens2023, Messa2022, Mestric2022}.  Together with the effect that the GMCs may not in coplanar, these conditions may directly produce stars with high velocity dispersion, naturally giving rise to thick-disk populations and providing a plausible explanation for the thick disks found in hydrodynamical simulations \citep{YuSijie2021, YuSijie2023}. Furthermore, the sloshing effect that perturbation from non-axisymmetric structures and the stochastic star formation in the disk induce bulk motion in the disk, increasing velocity dispersion of the stars \citep{Bland-Hawthorn2024, Bland-Hawthorn2025}.

(2) Thick disks were born thin but later thickened. The potential mechanisms include dynamical heating by minor mergers \citep{Quinn1993, Qu2011}, secular evolution by stellar structures \citep{Sellwood2014, Martinez-Medina2015}, and decreasing in surface mass density due to galaxy size growth.

\subsection{Thick disks begin as intermediate-scale-height, dynamically hot, dense structures}

The detection of relatively small disk scale heights at high redshifts was first reported by \citet{Elmegreen2017}, who found that the median scale height of disks at $z=3$ is $\sim0.3$\,kpc, which is although based on rest-frame $\sim0.2\,\micron$ observations with HST. This result was recently reinforced by \citet{Tsukui2025}, who analyzed JWST NIR rest-frame images of 111 edge-on disks. After correcting their measurements to a median stellar mass of $10^{9.2}\,M_\odot$, they obtained a median global disk scale height of 0.51\,kpc at $z=2.6$, compared to 0.77\,kpc at $z=0.26$. Consistently, using a JWST sample that is 15--30 times larger than in previous studies, we find a median scale height of $0.56$\,kpc at $z=3.25$. 
Together, these findings suggest that high-redshift disks are neither thin nor thick, but instead occupy an intermediate scale-height regime between the typical thin disks (0.41\,kpc for the Milky Way) and thick disks (1.23\,kpc for the Milky Way), and experienced vertical growth over Gyr timescales.  Consequently, the category~(1) mechanisms may involve physical conditions in their galaxy models that do not fully match real observations, and may therefore require further improvement.

This, however, does not exclude the role of gravitational-instability-driven scattering \citep{Bournaud2009} or the sloshing effect \citep{Bland-Hawthorn2025}. On the contrary, such processes likely play a crucial role in producing dynamically hot stellar populations within thick disks. In the same vein, the ``born thick'' disk stars formed from large-size GMCs or clumps, from bursty star formation seen in the simulations of \citet{YuSijie2023} also contribute to form hot stars. Our results are not necessarily inconsistent with these simulations, because the bridge connecting simulations and observations, namely, realistic mock observations derived from simulations, is still lacking.

Although disks at $z\approx3.5$ are geometrically intermediate-thickness rather than thin or thick, they are not scaled-up versions of local cold thin disks. Instead, they more closely resemble thick disks, as they are dynamically hot and exhibit high stellar mass surface densities, as discussed below. In Figure~\ref{fig:mu_mass}, we show that the stellar surface mass density, $\Sigma_{*, \rm disk}=M_{*,\rm disk}/(2\pi R_{e,\rm{disk},1\micron}^2)$, increases toward higher redshift: the median value is $\Sigma_{*, \rm disk}=10^{8.13}\,M_\odot\,\mathrm{kpc}^{-2}$ at $z=0.25$, and $\Sigma_{*, \rm disk}=10^{8.67}\,M_\odot\,\mathrm{kpc}^{-2}$ at $z=3.25$. A best-fit relation to the median values in the seven redshift bins yields
\begin{equation}\label{SigmaMass}
	\Sigma_{*, \rm disk}(z) = 10^{8.07\pm0.02} \exp\!\left((0.34\pm0.02)\,z\right) ~ M_\odot\,\mathrm{kpc}^{-2}.
\end{equation}
\noindent
Assuming an isothermal disk \citep{Kruit1981}, the vertical stellar velocity dispersion is related to $h_0$ and $\Sigma_{*, \rm disk}$ through (also see derivation in the Appendix):

\begin{equation}\label{Eq:h}
 h_0 = \frac{\sigma^2_{\rm *, vert}}{\pi\,G\,\Sigma_{*, \rm disk}},
\end{equation}

\noindent
or equivalently,


\begin{equation}
 \sigma_{\rm *, vert} = \sqrt{\pi\,G\,\Sigma_{*, \rm disk}\,h_0}.
\end{equation}

\noindent
Given $\Sigma_{*, \rm disk}=10^{8.67}\,M_\odot\,\mathrm{kpc}^{-2}$ and a bias-corrected $h_{0,1\micron}=0.56$\,kpc at $z=3.25$, we obtain $\sigma_{\rm *,vert}=59\,{\rm km\,s^{-1}}$. This value is consistent with $\sigma_{\rm *,vert}=50\pm5\,{\rm km\,s^{-1}}$ for stars at the solar neighborhood in the Milky Way’s thick disk \citep{Bland-Hawthorn2016}. If including the contribution of the gaseous component, assumed to account for 40\% of the total mass \citep{Narayanan2012}, the inferred velocity dispersion increases to $\sim78\,{\rm km\,s^{-1}}$.
Therefore, although the disks at $z=3.25$ appear geometrically thin, their vertical velocity dispersions of 59--78\,km\,s$^{-1}$ indicate that they are dynamically hot. These hot stellar populations are likely produced by star scattering \citep{Bournaud2009}, sloshing effect \citep{Bland-Hawthorn2025}, or formed inherently hot \citep{YuSijie2023}.  The strong gravitational potential of the dense mid-plane confines stars with high vertical velocity dispersions within a geometrically thin structure.

We may further gain insights from gas kinematics. Observations show that the velocity dispersion of the gaseous component, including the ionized, atomic, and molecular phases, increases systematically with redshift \citep{Glazebrook2013}. By compiling and analyzing measurements from the literature, \citet{Ubler2019} derived a redshift-dependent relation for the gas velocity dispersion: $\sigma_{\rm g}/{\rm km\,s}^{-1} = 10.9 + 11.0\,z$. Applying this relation yields $\sigma_{\rm g,vert}\sim50\,{\rm km\,s^{-1}}$ at $z=3.25$, consistent with the inferred $\sigma_{\rm *,vert}$ above. Such turbulent, gas-rich disks facilitate the formation of stars with high velocity dispersion \citep{YuSijie2023}. Consequently, even though the stellar disks at $z\approx3.5$ are geometrically intermediate-thickness rather than thick, with median $h_{0,1\micron}=0.56$\,kpc, they are dynamically hot.  Furthermore, these $z\approx3.5$ disks exhibit a single-component vertical structure and occupy a structural-parameter space comparable to that of thick disks identified at lower redshifts, while a distinct thin component only begins to emerge at $z<2$ \citep{Tsukui2025}. This may suggest that most galaxies initially form a thick disk, observed as a dense, dynamically hot, intermediate-thickness single component, followed by the subsequent development of a thin disk. 

Because our sample is selected to have an approximately constant median stellar mass across redshift, it does not trace the evolutionary paths of individual galaxies. However, \citet{Tsukui2025} similarly find relatively small scale heights ($\sim\,0.51$\,kpc) in high-redshift disks at a lower median mass of $10^{9.2}\,M_\odot$. Combined, these findings point toward a picture in which present-day thick disks originate from dense, dynamically hot, intermediate-thickness disks formed in the turbulent, gas-rich conditions of early cosmic epochs.

\subsection{Minor mergers and secular evolution do not drive present-day thick disks}

Dynamical heating by minor mergers \citep{Quinn1993, Qu2011} has long been proposed as a possible mechanism for thick disk formation, especially at early cosmic times when merger rates were significantly higher \citep{Duncan2019, OLeary2021}. This process, however, is predicted to induce strong vertical flaring of stellar disks \citep{Bournaud2009, Moster2010}, i.e., an intrinsic increase of scale height with radius. Yet, observations suggest otherwise. \cite{Bournaud2009} showed that high-redshift disks do not exhibit pronounced flaring, at least based on visual inspection of edge-on images. \cite{Elmegreen2006} did not find signature of minor mergers in their sample. Moreover, \cite{Bizyaev2014} pointed out that even a slight deviation from a perfectly edge-on orientation can produce artificial flaring, which may mimic the expected intrinsic trend.

As shown in Figure~\ref{fig:rad_grad}, the observed radial gradient of the measured scale height in our sample is consistent with the amount of artificial flaring expected from the projection effect and the residual PSF effect (see Section~\ref{artificial_grad} for detailed calculation). This agreement implies that the apparent flaring in our sample is likely due to observational effects rather than intrinsic structural variations. Therefore, we find no evidence for intrinsic vertical flaring across our edge-on disk sample over the redshift range $0<z<3.5$, effectively ruling out minor mergers as the dominant mechanism responsible for the formation of the thick disk component.

Nevertheless, mergers likely played an important role in specific cases such as the Milky Way, where they contributed to disk heating and the formation of in-situ halo stars with chemical abundances similar to those of thick disk stars \citep{Bonaca2020, Belokurov2020, Conroy2022}.

Our results also do not support scenarios in which thick disks formed primarily through slow heating of a pre-existing thin disk by bar buckling \citep{Sellwood2014} or spiral scattering \citep{Martinez-Medina2015}. These mechanisms are potentially important, as bars and spirals are common in nearby galaxies \citep{Yu2018, YuHo2018, Yu2019, YuHo2020} and have been shown to play a key role in driving galaxy secular evolution \citep{Yu2021, Yu2022b, Yu2022}. If bar buckling or spiral scattering played a dominant role in disk thickening, one would expect disk thickness to increase toward lower redshifts at $z < 1$, when bars and spiral arms are well developed and dynamically active. In contrast, we observe a clear decrease in disk thickness over this redshift range (Fig.~\ref{fig:hz_comp_z}). Furthermore, these secular mechanisms are also disfavored by the observed high-[$\alpha$/Fe] stars in the thick disk of the Milky Way. If they were responsible for building the thick disk over the last 8\,Gyr of the Universe, the gas reservoir would inevitably produce a substantial number of low-[$\alpha$/Fe] stars that would be heated into the thick disk, which is contrary to observations. Consistently, hydrodynamical simulations also show no significant thick disk formation at $z\lesssim1$, after the end of the bursty star-formation phase \citep{YuSijie2023}.

\subsection{Scale heights from $z=3$ to 1}\label{SH31}

From $z=3.25$ to $z=1.25$, the disk scale height increases from $0.56$ to $0.84$\,kpc and $R_d$ grows from $1.52$ to $2.56$\,kpc, while $\Sigma_{*, \rm disk}$ declines from $10^{8.67}$ to $10^{8.24}\,M_\odot\,\mathrm{kpc}^{-2}$. The errors of these values are approximately $0.015$\,kpc, $0.05$\,kpc, and $10^{7.3}\,M_\odot\,\mathrm{kpc}^{-2}$, respectively. For comparison, $\Sigma_{*, \rm disk}=10^{8.07}\,M_\odot\,\mathrm{kpc}^{-2}$ at $z=0$ in our sample. As discussed below, these trends reflect a combination of effects, including the decreasing of $\Sigma_{*, \rm disk}$ due to size growth, and the generation of dynamically hot stars formed through bursty star formation, stellar scattering, and the sloshing effect.

We first examine the impact of size growth. It is well established that, at fixed stellar mass, galaxy size increases toward lower redshift \citep{vanderWel2014, Allen2025, Yang2025}, likely as a result of dark-matter halo growth and radial stellar migration. Because our galaxies have a nearly constant stellar mass of $10^{10.4}\,M_\odot$ across redshift, size growth naturally leads to a lower $\Sigma_{*, \rm disk}$, which plays a key role in determining disk scale height. To test whether the decreasing of $\Sigma_{*, \rm disk}$ due to size growth alone can account for the observed increase in $h_0$, we assume that there are no exchanges of mass, energy, or angular momentum between the system and its environment to isolate the pure effect of size growth.  The size growth is therefore assumed to be adiabatic.  The scale height varies with $\Sigma_{*, \rm disk}$ as follows (See Appendix for formula derivation):
\begin{equation}
  h_0 \propto \Sigma_{*, \rm disk}^{-0.2}.
\end{equation}
\noindent
This relation predicts an increase in scale height by a factor of $1.16$ from $z=3.25$ to $1.75$, smaller than the observed factor of $1.4$. Hence, additional energy sources are required to raise the stellar velocity dispersion.

Both observations and simulations suggest that thick disks were established rapidly during a turbulent phase rather than through a gradual buildup at early cosmic times ($z\gtrsim1$) \citep{Nissen2020, Xiang2022, Conroy2022, YuSijie2021, YuSijie2023, xiang2025}. During this phase, fueled by intense gas inflows along cosmological cold streams \citep{Ceverino2010}, high-redshift disks exhibited elevated star formation rates \citep{Wuyts2011, Speagle2014, Gillman2024, Morishita2024} and clumpy, irregular morphologies \citep{Conselice2008, Mortlock2013, Ferreira2022b, Kartaltepe2023, Faisst2025}, driven by violent gravitational instabilities \citep{Romeo2010, Romeo2014}. In such physical conditions, dynamically hot stars that form the thick disk may form, via star scattering \citep{Bournaud2009} or  the sloshing effect that perturbation from non-axisymmetric structures and  stochastic star formation in the disk induce bulk motion in the disk respect to the halo potential, increasing the velocity dispersion \citep{Bland-Hawthorn2024, Bland-Hawthorn2025}.

If we consider an extreme case where bursty star formation and gravitational instabilities are strong enough to keep $\sigma_{\rm *,vert}$ unchanged during size growth (isothermal thickening), the scale height follows (see Eq.~(\ref{Eq:h}))
\begin{equation}
h_0 \propto \Sigma_{*, \rm disk}^{-1}.
\end{equation}
This predicts an increase in scale height by a factor of $2.1$ from $z=3.25$ to $1.75$, larger than observed. Thus, the actual evolution lies between the adiabatic and isothermal limits. The observed rise in scale height at high redshift therefore reflects the combined effects of size growth, bursty star formation, and violent disk instabilities. These disks are single-component systems that serve as progenitors of present-day thick disks, followed by the subsequent development of a thin disk.

After $z\approx1$--2, the penetration of cold streams into massive halos becomes inefficient as halos develop shock-heated gas atmospheres, leading to a decline in gas accretion rates \citep{Dekel&Birnboim2006, Dekel2009Natur, Ceverino2010}. Consequently, galaxies become less clumpy, reducing stellar scattering and the rate of thickening.  Simulations have demonstrated that at this epoch some stars with low vertical velocity dispersion begin to form \citep{YuSijie2023} facilitating the formation of old thin disks.

Observations have also confirmed the presence of old thin disks \citep{Comeron2016, Elmegreen2017, Tsukui2025}. \cite{Comeron2016} found old stellar populations in thin disks of S0 galaxies, and \cite{Elmegreen2017} inferred thin disks at $z\sim2$ from vertical color gradients. Using JWST imaging, \cite{Tsukui2025} decomposed disks at $z>1$ and identified the first thin disk at $z=1.96$, finding that the thin-disk fraction increases toward lower redshift. The detection of spiral arms and bars in high-redshift galaxies \citep{Kartaltepe2023, Ferreira2022b, Liang2024, LeConte2024, XuYu2024, LeConte2025, Yu2025, Huertas-Company2025, Guo2025, Geron2025} further supports the presence of dynamically cold, settled disks \citep{Kraljic2012, ElmegreenElmegreen2014}. \cite{Tsukui2025} show that the thin disk forms in a downsizing way: thin disks appear to form earlier in more massive galaxies, and on average thin disks appear to dominate over the past 8--9\,Gyr for the stellar mass bin $2.5\times10^{10}\,M_\odot$, corresponding to the bend in Figure~\ref{fig:hz_comp_z}.

\subsection{Scale heights from $z=1$ to 0}\label{SH10}

At $z\approx1$, the virialization of the inner circumgalactic medium begins to regulate cosmological cold streams, reducing gas inflow to the disk and marking the transition from the bursty phase to a steady star-forming phase that enables the formation of dynamically cold stellar disks \citep{Ceverino2010, YuSijie2023}. Although minor mergers may still occur during this later phase at $z\lesssim1$, simulations suggests that they do not account for the majority of thick-disk stars \citep{YuSijie2023}, consistent with our result that no intrinsic flaring in the disks is observed. Instead, minor mergers mildly heat some thin-disk stars and contribute to the young-star tail of the thick-disk population \citep{YuSijie2023}. Fueled by cold gas accretion into the mid-plane and leftover gas from previous phase, the thin disk gradually builds up through the steady star formation.

Toward $z=0$, galaxy sizes at fixed stellar mass continue to increase \citep{vanderWel2014}, reducing $\Sigma_{*, \rm disk}$ and the mid-plane gravitational potential, thereby tending to increase the scale height of the disk. This effect competes with thin-disk formation, which drives a reduction in the global scale height through two mechanisms. First, the deepening of the mid-plane gravitational potential by cold gas accretion contracts the thick disk vertically \citep{ElmegreenElmegreen2006}. These effects become more pronounced with the progressive growth of the thin disk over cosmic time, driving the overall decline in global thickness. Second, the superposition of thin and thick components introduces a geometric effect that reduces the measured global scale height in single-disk sech$^2$ fitting of the combined vertical profile.  During this period, Type Ia supernovae enrich the interstellar medium with Fe, lowering the [$\alpha$/Fe] ratios characteristic of thin-disk stars \citep{Bland-Hawthorn2016}.

The ratio of scale length to scale height ($R_{d,1\micron}/h_{0,1\micron}$), i.e., the inverse of the relative scale height, serves as a useful indicator of the galaxy’s dynamical state, roughly tracing the $v/\sigma$ of the system \citep{Kormendy1982}. The bias-corrected values are summarized in Table~\ref{tab:results} and shown in Figure~\ref{fig:LH}. The ratio remains approximately constant at $2.7\pm0.2$ for $z>1.5$, but increases to $4.0\pm0.4$ at $z=0.25$, indicating that galaxies become increasingly rotationally supported toward lower redshift, which is consistent with the picture in which thin-disk growth dominates galaxy evolution in this epoch.

\subsection{Simplified Model for Evolution and Its Fitting}\label{model}

Based on the discussion above, we construct a simplified model for the redshift evolution of scale height, aiming to gain further insight on the galaxy disk evolution (Figure~\ref{fig:evo}).  Since the mid-plane mass density is the key factor determining disk thickness, the model is formulated as a function of $\Sigma_{*, \rm disk}$, which evolves with redshift and is connected back to the redshift according to the best-fit relation in Eq.~(\ref{SigmaMass}).  We begin by assuming the dense, dynamically hot, intermediate-thicness disk observed at $z\approx3.5$ represent the progenitors, or ``seed galaxies'', of present-day thick disks. An example of its vertical density profile is shown in the bottom-right inset panel of Figure~\ref{fig:evo}.  As previously demonstrated, the scale-height increasing, due to the decreasing in $\Sigma_{*, \rm disk}$, violent gravitational instabilities and bursty star formation, between $z=3.5$ and $z=1$, is a intermediate process between the adiabatic process and isothermal regimes, we parameterize the relation between scale height and $\Sigma_{*, \rm disk}$ as  
\begin{equation}\label{medadia}
  h_0 = D\,\Sigma_{*, \rm disk}^{k},
\end{equation}
\noindent
where $D$ is a normalization constant and $k$ lies between $-0.2$ and $-1$. Here we adopt kpc and $M_{\odot}\,{\rm kpc}^{-2}$ as the unit of scale height and stellar mass surface density, respectively. For simplicity, the increase in the mass of gas component is absorbed into the factor $D$, given that both the stellar surface density $\Sigma_{*, \rm disk}$ and the gas fraction increase with redshift.

For $z\lesssim 1$, the formation of thick disks are expected to significantly reduced. We therefore assume pure adiabatic thickening for the thick disk at $z<1$, and use a step function $\zeta$ to connect $z>1$ and $z<1$ formulae (before addition of the thin disk):
\begin{equation}\label{thick_ori}
\begin{split}
  h_{\rm thick}^{0}(z) = D\biggl[
      \zeta(z)\,\Sigma_{*,{\rm disk}}^{\,k+0.2}(z=1)\,
        \Sigma_{*,{\rm disk}}^{-0.2}(z)  \\
    + \bigl(1 - \zeta(z)\bigr)\,\Sigma_{*,{\rm disk}}^{\,k}(z)
  \biggr].
\end{split}
\end{equation}
\noindent
where $\Sigma_{*,{\rm disk}}^{k+0.2}(z=1)$ is to ensure the continuity at $z=1$.   Motivated by the detection of a thin disk as early as $z\sim1.96$ and the increasing number fraction of thin disks toward lower redshift reported in \cite{Tsukui2025}, we adopt the following functional form for the mass fraction of thin disk:
\begin{equation}\label{fthin}
  f_{\rm thin} = 
\left\{
\begin{array}{ll}
f_0\, (2-z)^{t}, & \text{for } z \leq 2 \\
0, & \text{for } z > 2
\end{array}
\right.
\end{equation}
\noindent
where $f_0$ and $t$ characterize the amplitude and growth rate. We assume the mass-to-light ratio for the thick and thin disks is 1.2/1 \citep{Comeron2011}. \cite{Tsukui2025} find that scale heights for thin disk are approximately unchanged with $z$ with a median value of $\sim\,0.33\,$kpc. We therefore adopt a constant scale height for the thin-disk model: 
\begin{equation}\label{thin_ori}
  h_{\rm thin}(z) = 0.43\,{\rm kpc}
\end{equation}
\noindent
scaled upward by a factor of 1.3, difference between their and our global scale heights, to account for stellar mass differences between their sample and ours. This value is almost the same with the sech$^2$ scale height of the Milky Way's thin disk (0.41\,kpc). An example vertical light profile showing a thick disk overlapped with a newly formed thin disk is displayed in the inset at $z=1.5$, while a later stage with a more developed thin disk is illustrated at $z=1$ in Figure~\ref{fig:evo}.

Because our sample has constant stellar mass across redshift, the growth of the thin disk must be accompanied by a reduction of stellar mass in the thick component. To model this redistribution, we assume that the vertical velocity dispersion of thick-disk stars remains unchanged, with a modest increase caused by adiabatic contraction during thin-disk buildup. This is reasonable because the dominant heating mechanisms responsible for thick-disk formation operate primarily at $z\gtrsim1$ and are largely ineffective at later times. The gradual buildup of the thin disk, with surface density $f_{\rm thin}\Sigma_{*, \rm disk}(z)$, is assumed to proceed adiabatically, contracting the vertical structure and slightly increasing the velocity dispersion of stars of the pre-existing thick disk \citep{Elmegreen2006, Elmegreen2017}.  We assume the approximation that the thin and thick components each maintain independent pressure equilibria, while their gravitational potentials are coupled \citep{Elmegreen2006, Jog2007}. 
The connection between the scale heights of the thin and thick components in a thick-thin system can then be approximated as (see Appendix and \citealt{Forbes2012}):
\begin{equation}\label{thick_response}
h_{\rm thick} = \frac{\sigma_{\rm thick}^2}{\pi G \left( (1-f_{\rm thin})\Sigma_{*, \rm disk} + f_{\rm thin}\Sigma_{*, \rm disk} \dfrac{h_{\rm thick}}{h_{\rm thin}} \right)},
\end{equation}
where $\sigma_{\rm thick}$ is linked to $h_{\rm thick}^0$ and $\Sigma_{*, \rm disk}$ through the adiabatic invariant of the thick disk (Eq.~\ref{adiabatic_invar}).
The quantity $h_{\rm thick}(z)$ is then solved numerically.

The corresponding vertical light profiles are given by
\begin{equation}
  I_{\rm thick}(h,z) = \frac{(1-f_{\rm thin})\,\Sigma_{*, \rm disk}(z)}{2\,h_{\rm thick}(z)}\,{\rm sech}^2\!\left( \frac{h}{h_{\rm thick}(z)}\right)
\end{equation}
for the thick disk and 
\begin{equation}
  I_{\rm thin}(h, z) =
  \frac{f_{\rm thin}\,\Sigma_{*,\rm disk}(z)}{2\,h_{\rm thin}(z)}
  \,{\rm sech}^2\!\left(\frac{h}{h_{\rm thin}(z)}\right)
\end{equation}
\noindent
for the thin disk. An example of the combined thin+thick profile at $z=0$, showing a dominant thin disk embedded within a thicker component, is presented in Figure~\ref{fig:evo}. To obtain the disk scale height averaging the thin and thick disk, we fit a single sech$^2$ function to the combined profile, mimicking our observational procedure applied to edge-on disks in COSMOS-Web:
\begin{equation}\label{final_h}
h_0(z) =
\Gamma\!\left(\,I_{\rm thin}(h, z) + I_{\rm thick}(h, z)\,\right),
\end{equation}
where $\Gamma$ denotes the fitting operation.

We fit Eq.~(\ref{final_h}) to the bias-corrected $h_{0,1\micron}(z)$ data using four free parameters: $D$, $k$, $f_0$, and $t$. The best-fit function is shown as the black curve in Figure~\ref{fig:hz_comp_z} and \ref{fig:evo}. To estimate uncertainties, we perform 1000 bootstrap resamplings with replacement and refit the model each time; the resulting best-fit curves are shown as gray lines in Figure~\ref{fig:hz_comp_z} and \ref{fig:evo}. The resulting parameters yield $D=10^{6.0}\pm10^{4.9}$, $k=-0.73\pm0.02$, $f_0=0.11\pm0.03$, and $t=2.52\pm0.52$. The best-fit evolutionary curve for the thick-disk component is plotted as a red dashed line in Figure~\ref{fig:evo}. The continued increasing of the thick-disk scale height below $z=1$ is primarily due to the decline in $\Sigma_{*, \rm disk}$ caused by size growth and mass reduction. The redshift corresponding to the peak of the best-fit curve is $z_{\rm peak}=1.12\pm0.04$.

We assess the robustness of the fit by comparing the Bayesian Information Criterion (BIC) between Eq.~(\ref{final_h}) and a simple linear model. The BIC for Eq.~\ref{final_h} is lower by 20, indicating a strong statistical preference for our simplified model and confirming the presence of a peak in the evolution of disk scale height.

The best-fit thin-disk mass fraction, $f_{\rm thin}$, is shown in Figure~\ref{fig:f_thin}. It increases from 0 at $z=2$ to $0.10\pm0.03$ at $z=1$ and $0.60\pm0.10$ at $z=0$. Our inferred $f_{\rm thin}$ values are systematically higher than those reported by \citet{Tsukui2025} for a sample with a median mass of $10^{9.2}\,M_\odot$, likely because galaxies with higher stellar masses tend to host a larger fraction of thin disks (\citealt{Comeron2018} and see Figure~10 of \citealt{Tsukui2025}).  Another plausible explanation is that the decomposition of thin and thick disk components may be systematically incomplete relative to the idealized model expectations. Detecting two different components requires sufficient contrast between their vertical profiles, so galaxies with comparable thin and thick disk masses are more likely identified as having a two-component structure. In contrast, when either component dominates the flux, the secondary component becomes difficult to detect observationally and the galaxy is more likely classified as a single-component system. This may lead to the absence of very low thin-disk fractions in massive galaxies reported by \cite{Tsukui2025} and \cite{Comeron2018}. Nevertheless, the inferred fraction at $z=0$ agrees well with the measurements by \cite{Comeron2018}, which gives $f_{\rm thin}=0.66\pm0.09$ for nearby galaxies with $M_*\sim10^{10}$--$10^{10.8}\,M_\odot$. This agreement suggests that, despite its simplicity, our model captures the essential physical processes governing the coevolution of thin and thick disks.

In summary, the observed redshift evolution of disk scale height from COSMOS-Web can be explained by the interplay between thick- and thin-disk formation and evolution. The rise in $h_0$ from $z=3.5$ to $1$ results from size-driven decreasing of $\Sigma_{*, \rm disk}$, violent gravitational scattering, and bursty star formation. The subsequent decline from $z=1$ to $0$ reflects geometric overlap with thin disks, continuing $\Sigma_{*, \rm disk}$ decreasing, and contraction driven by thin-disk growth.

\section{Summary and Conclusion} \label{sec:summary}

We conducted a comprehensive analysis of the global disk scale heights for a sample of 2631 clean, nearly edge-on disk galaxies at $0<z<3.5$, observed by JWST/NIRCam in the COSMOS-Web survey with the F115W, F150W, F277W, and F444W filters. The sample has a median stellar mass of $2.5\times10^{10}\,M_{\odot}$, and a disk-axis ratio cut of $q_d < 0.4$ was applied to ensure edge-on orientations.  We measured the vertical scale height, $h_0$, by fitting a 2D model consisting of a bulge and an edge-on disk described by a single sech$^2$ vertical profile. The disk size is characterized by the radial exponential scale length $R_d$. The resulting $h_0$ represents an average scale height that combines the contributions of both thin and thick disks, where such components coexist. For comparison, exponential vertical profiles were also tested. In addition, 1D vertical light profiles were fitted with a line spread function (LSF), constructed using a PSF, to derive scale heights as a function of radius. The measurements of scale heights and scale lengths, as well as the correction curves used to account for projection effects, can be accessed at Zenodo \citep{Yu2026Zenodo}. Our main findings are summarized as follows:

\begin{enumerate}
  \item $h_0$ derived from sech$^2$ vertical profiles are 1.37 times as large as those obtained from exponential profiles, with negligible scatter.
  
  \item $h_0$ measured at shorter rest-frame wavelengths are systematically overestimated relative to those measured at longer wavelengths, primarily due to dust extinction across the mid-plane. This effect is consistent with \citet{Bizyaev2009}. We adopt a fixed rest-frame 1\,\micron\ for our measurements to mitigate the dust extinction and spatial variation in stellar population.   
  
  \item Deviations from a perfectly edge-on inclination systematically bias $h_0$ toward larger values, with this projection effect becoming progressively more severe in relatively thinner disks (i.e., those with smaller $h_0/R_d$). We developed a new statistical correction method for this inclination bias, which is more accurate than the constant correction factor proposed by \citet{HamiltonCampos2023}.
  
  \item The observed radial variation of scale heights across our sample is consistent with artificial flaring primarily caused by projection effects, with a minor contribution from imperfect PSF treatment in the 1D fitting. This disfavors minor mergers as the dominant mechanism for the disk thickening and hence thick-disk formation, which are expected to produce genuine flaring.
  
  \item After correcting for the biases caused by variation in observed rest-frame wavelength and by the projection effect,
   we find an unambiguous trend in scale height as a function of redshift: it increases from $0.56\pm0.03$\,kpc at $z=3.25$ to $0.84\pm0.04$\,kpc at $z=1.25$, and then decreases to $0.67\pm0.06$\,kpc at $z=0.25$.  The scatters among individual measurements are 0.2--0.3\,kpc.  The disks at $z\sim3.5$ were geometrically intermediate-thickness, yet dynamically hot and highly dense, likely representing the progenitors of present-day thick disks.

  \item The bias-corrected ratio of scale length $R_d$ to $h_0$ ($R_d/h_0$) remains approximately constant at $2.7\pm0.2$ at $z>1.5$, while increases to $4.0\pm0.4$ at $z=0.25$.
  
  \item Using a simplified evolutionary model, we suggest that the early dense, dynamically hot, intermediate-thickness, single-disk galaxies at $z\approx3.5$ thickened as a result of decreasing surface mass density (as galaxies grew in size), violent gravitational instabilities, and bursty star formation, leading to increasing scale height toward lower redshift at $z>1$. Below $z\sim1$, thin-disk growth became dominant, producing a vertically more compact structure with smaller scale heights toward $z=0$.

  \item The model fitting further indicates that the thin-disk mass fraction $f_{\rm thin}$ increases from $f_{\rm thin}=0$ at $z\approx3.5$, to $f_{\rm thin}=0.10\pm0.03$ at $z=1$, and to $f_{\rm thin}=0.60\pm0.10$ at $z=0$. The best-fit value at $z=0$ agrees well with the near-infrared observational constraints for nearby galaxies with comparable stellar mass reported by \citet{Comeron2018}.
\end{enumerate}

A detailed investigation of disk thickness as a function of galaxy properties, such as stellar mass, will be pursued in future work. Future studies focusing on large-sample thick-thin disk decompositions, as well as direct comparisons between realistic mock images from high-resolution hydrodynamical simulations and observations, will be essential for disentangling the physical processes that govern the evolution of disk thickness and the transition from high-redshift dense hot disks with intermediate-scale-height to present-day thin–thick disk systems.

\begin{acknowledgments}
We thank the referee for the insightful comments and suggestions.
LCH was supported by the China Manned Space Program (CMS-CSST-2025-A09) and the National Science Foundation of China (12233001). TT is supported by the JSPS Grant-in-Aid for Research Activity Start-up (25K23392) and the JSPS Core-to-Core Program (JPJSCCA20210003). SYU acknowledges support from the UTokyo Global Activity Support Program for Young Researchers. We thank the discussion with Jing Wang, Jianhui Lian, Raymond C. Simons, Qikang Feng, and Sijia Li. Kavli IPMU is supported by World Premier International Research Center Initiative (WPI), MEXT, Japan. The JWST data presented in this article were obtained from the Mikulski Archive for Space Telescopes (MAST) at the Space Telescope Science Institute. The specific observations analyzed can be accessed via \dataset[DOI: 10.17909/ph8h-qf05]{https://doi.org/10.17909/ph8h-qf05}. 
\end{acknowledgments}

\begin{contribution}
SYU and LCH developed the initial idea of the project. SYU conducted the data reduction, developed the methods, and performed analysis. SYU wrote the manuscript, with revisions from other contributors.

\end{contribution}

%
\facilities{JWST(NIRCam) }

\software{astropy \citep{Astropy2013,Astropy2018,Astropy2022},  
          IMFIT \citep{Erwin2015}, 
          SEP \citep{Bertin1996, Barbary2016} }

\appendix
 
\section{Derivation of Key Equations for a Self-Gravitating Single-Disk Model}
In the main text, we use $h$ to denote the vertical distance instead of the commonly used $z$, to avoid confusion with the symbol for the redshift. In the Appendix, however, we revert to $z$ for vertical distance, as it is the convention most widely adopted in the literature.

For a stellar disk in vertical equilibrium, the balance between the vertical component of the gravitational force and the dynamical pressure ($P=\rho\,\sigma^2$) associated with the random stellar motions can be expressed as
\begin{equation}
    \frac{1}{\rho}\frac{d(\rho\,\sigma_z^2)}{dz} = -\frac{d\Phi}{dz},
\end{equation}
where $\rho(z)$ is the mass density, $\sigma_z$ is the vertical velocity dispersion, and $\Phi(z)$ is the gravitational potential. This is the vertical component of the steady-state Jeans equation for a collisionless stellar system. If the system is isothermal in the sense that $\sigma_z$ is constant with height, the equation reduces to
\begin{equation}
    \sigma_z^2\,\frac{d\rho}{dz} = -\rho\,\frac{d\Phi}{dz}.
\end{equation}
The gravitational potential obeys the Poisson equation
\begin{equation}
    \frac{d^2\Phi}{dz^2} = 4\pi G\rho.
\end{equation}
Combining these two equations and eliminating $\Phi$ gives
\begin{equation}\label{A4}
    \frac{d^2}{dz^2}\ln\rho = -\frac{4\pi G}{\sigma_z^2}\rho,
\end{equation}

\noindent
whose analytic solution is

\begin{equation}
    \rho(z) = \rho_0\,\mathrm{sech}^2\!\left(\frac{z}{z_0}\right),
\end{equation}
where $\rho_0$ is the midplane density and $z_0$ is the vertical scale height.
Substituting this form into Eq.~(\ref{A4}) yields
\begin{equation}\label{A6}
    z_0^2 = \frac{\sigma_z^2}{ 2\pi G\rho_0}.
\end{equation}


The total surface density of the disk is
\begin{equation}
    \Sigma = \int_{-\infty}^{+\infty} \rho(z)\,dz = 2\rho_0 z_0.
\end{equation}
Substituting $\rho_0 = \Sigma / (2z_0)$ into Eq.~(\ref{A6}) gives
\begin{equation}\label{eq:sigma_relation}
    z_0 = \frac{\sigma_z^2}{\pi G \Sigma}.
\end{equation}

If the disk evolves slowly in an adiabatic manner, the vertical pressure and density satisfy
\begin{equation}
    P = K\rho^{\gamma},
\end{equation}
where $K$ is the adiabatic constant and $\gamma=5/3$ for a collisionless stellar system. With $P_0=\rho_0\sigma_z^2$ and $\rho_0=\Sigma/(2z_0)$, we obtain
\begin{equation}\label{adiabatic_invar}
    \sigma_z^2 = K\left(\frac{\Sigma}{2z_0}\right)^{\gamma-1}.
\end{equation}
Combining this with Equation~\ref{eq:sigma_relation} yields
\begin{equation}
    \Sigma^{\,2-\gamma}z_0^{\,\gamma}=\mathrm{const.}
\end{equation}
For $\gamma=5/3$, the scale height and velocity dispersion vary as
\begin{equation}
    z_0 \propto \Sigma^{-1/5}, \qquad
    \sigma_z \propto \Sigma^{2/5}.
\end{equation}
Thus, during a slow adiabatic expansion, the disk becomes slightly thicker and dynamically cooler as its surface density decreases.

\section{Derivation of Key Equations for a Self-Gravitating Double-Disk Model}

To derive an approximate formula that connects the scale heights of the two disk components, we assume that the thin and thick components each maintain independent pressure equilibria, while their gravitational potentials are coupled. We emphasize that, within this approximation, the two resulting sech$^2$ profiles do not constitute a strictly self-consistent solution to the Poisson equation together with vertical hydrostatic balance. Even so, this approximation and the adoption of two sech$^2$ profiles offer a practical and decent representation of the observed scale heights. 
Each disk follows a sech$^2$ density profile:

\begin{equation}
\rho_i(z) = \rho_{0i} \operatorname{sech}^2\left(\frac{z}{z_{0i}}\right), \quad
i = 1,2,
\end{equation}

with corresponding surface densities

\begin{equation}
\Sigma_i = 2\rho_{0i} z_{0i}.
\end{equation}

The total gravitational potential satisfies the Poisson equation:

\begin{equation}
\frac{d^2 \Phi}{dz^2} = 4\pi G \left[ \rho_1(z) + \rho_2(z) \right].
\end{equation}

Each disk is assumed to be isothermal in the vertical direction, so that vertical hydrostatic equilibrium for each component gives

\begin{equation}
\sigma_i^2 \frac{d \ln \rho_i}{dz} = - \frac{d\Phi}{dz},
\end{equation}

where \(\sigma_i\) is the vertical velocity dispersion of disk \(i\).

Taking the logarithmic derivative of the sech$^2$ profile, we obtain

\begin{equation}
\frac{d \ln \rho_i}{dz} = - \frac{2}{z_{0i}} \tanh\left( \frac{z}{z_{0i}} \right),
\end{equation}

so that hydrostatic equilibrium becomes

\begin{equation}
\frac{d\Phi}{dz} = \frac{2\sigma_i^2}{z_{0i}} \tanh\left( \frac{z}{z_{0i}} \right).
\end{equation}

Differentiating once more with respect to \(z\) gives

\begin{equation}
\frac{d^2\Phi}{dz^2} = \frac{2\sigma_i^2}{z_{0i}^2} \operatorname{sech}^2\left(\frac{z}{z_{0i}}\right).
\end{equation}

Evaluating at the mid-plane \(z=0\) where sech$^2(0)=1$ and comparing with the Poisson equation yields

\begin{equation}
\frac{2\sigma_1^2}{z_{01}^2} = \frac{2\sigma_2^2}{z_{02}^2} = 4 \pi G (\rho_{01} + \rho_{02}).
\end{equation}

Substituting \(\rho_{0i} = \Sigma_i / (2 z_{0i})\) gives the coupled algebraic equations for the scale heights:

\begin{equation}
\sigma_1^2 = \pi G \left( \Sigma_1 z_{01} + \Sigma_2 \frac{z_{01}^2}{z_{02}} \right),
\quad
\sigma_2^2 = \pi G \left( \Sigma_2 z_{02} + \Sigma_1 \frac{z_{02}^2}{z_{01}} \right).
\end{equation}

Dividing the two equations provides a simple relation between the scale heights:

\begin{equation}
\frac{z_{01}}{z_{02}} = \frac{\sigma_1}{\sigma_2}.
\end{equation}

The explicit expressions for \(z_{01}\) and \(z_{02}\) are then

\begin{equation}
\label{eq:z01z02}
z_{01} = \frac{\sigma_1^2}{\pi G \left( \Sigma_1 + \Sigma_2 \dfrac{z_{01}}{z_{02}} \right)}, \quad
z_{02} = \frac{\sigma_2^2}{\pi G \left( \Sigma_2 + \Sigma_1 \dfrac{z_{02}}{z_{01}} \right)}.
\end{equation}


\end{document}